\documentclass[final,5p,times,twocolumn,authoryear]{elsarticle}

\usepackage[utf8]{inputenc}
\usepackage{textgreek}
\usepackage{textcomp} 

\usepackage{amssymb}
\usepackage{amsmath}

\usepackage{graphicx} 
\graphicspath{ {figures/} }

\usepackage{caption}
\usepackage{subcaption}

\usepackage{hyperref}

\usepackage{xcolor}

\usepackage[capitalise, nameinlink]{cleveref}

\crefname{chapter}{Chap.}{Chaps.}
\crefname{section}{Sect.}{Sects.}
\crefname{figure}{Fig.}{Figs.}
\Crefname{chapter}{Chapter}{Chapters}
\Crefname{section}{Section}{Sections}
\Crefname{figure}{Figure}{Figures}

\journal{Astronomy $\&$ Computing}

\begin{document}

\begin{frontmatter}



\title{Towards an optimal extraction of cosmological parameters from galaxy cluster surveys using convolutional neural networks}


\author[unimi,infn-milan]{I. Sáez-Casares}
\author[oavda]{M. Calabrese}
\author[unimi,inaf-brera]{D. Bianchi}
\author[lapth]{M. S. Cagliari}
\author[unimi]{M. Chiarenza}
\author[oavda]{J.-M. Christille}
\author[unimi,inaf-brera]{L. Guzzo}

\affiliation[unimi]{
            organization={Dipartimento di Fisica `Aldo Pontremoli', Università degli Studi di Milano},  
            addressline={Via Celoria 16}, 
            city={I-20133 Milan},
            country={Italy}
}
\affiliation[infn-milan]{
            organization={INFN, Sezione di Milano},
            addressline={Via Celoria 16},
            city={I-20133 Milan},
            country={Italy}
}
\affiliation[oavda]{
            organization={Astronomical Observatory of the Autonomous Region of Aosta Valley (OAVdA), Fondazione Clément Fillietroz - ONLUS},
            addressline={Loc. Lignan, 39},
            city={I-11020 Nus},
            country={Italy}
}
\affiliation[inaf-brera]{
            organization={INAF, Osservatorio Astronomico di Brera},
            addressline={via Brera 28},
            city={I-20121 Milan},
            country={Italy}
}
\affiliation[lapth]{
            organization={Laboratoire d'Annecy de Physique Theorique (LAPTh), CNRS/USMB},
            addressline={99 Chemin de Bellevue BP110 - Annecy -  F-74941 - ANNECY CEDEX - FRANCE},
}

\begin{abstract}

The possibility to constrain cosmological parameters from galaxy surveys using field-level machine learning methods that bypass traditional summary statistics analyses, depends crucially on our ability to generate simulated training sets.
The latter need to be both realistic, as to reproduce the key features of the real data, and produced in large numbers, as to allow us to refine the precision of the training process.
The analysis presented in this paper is an attempt to respond to these needs by (a) using clusters of galaxies as tracers of large-scale structure, together with (b) adopting a 3LPT code (\texttt{Pinocchio}) to generate a large training set of $32\,768$ mock X-ray cluster catalogues.
X-ray luminosities are stochastically assigned to dark matter haloes using an empirical $M-L_X$ scaling relation. 
Using this training set, we test the ability and performances of a 3D convolutional neural network (CNN) to predict the cosmological parameters, based on an input overdensity field derived from the cluster distribution.
We perform a comparison with a neural network trained on traditional summary statistics, that is, the abundance of clusters and their power spectrum.
Our results show that the field-level analysis combined with the cluster abundance yields a mean absolute relative error on the predicted values of $\Omega_{\rm m}$ and $\sigma_8$ that is a factor of $\sim 10 \%$ and $\sim 20\%$ better than that obtained from the summary statistics. Furthermore, when information about the individual luminosity of each cluster is passed to the CNN, the gain in precision exceeds $50\%$.  
\end{abstract}




\end{frontmatter}




    
    
    

\section{Introduction}
\label{sec:intro}

Large surveys of extragalactic objects, used as tracers of the large-scale distribution of matter, are a cornerstone of the standard cosmological model, usually known as $\Lambda$CDM.
The latter represents one of the major scientific achievements of the twentieth century, yet the very nature of its fundamental components remains unknown \citep[see e.g.][]{Amendola_2018}.
In this model, about 25\% of the mass--energy content of the Universe consists of {\it cold dark matter} (CDM), which should be composed by one or more species of massive particles, for which there is still no direct evidence  \citep[see e.g.][]{Battaglieri_2017}.
Additionally, a 70\% contribution from {\it dark energy}, apparently in the form of a nonzero cosmological constant $\Lambda$, is required to explain the acceleration of cosmic expansion  discovered less than three decades ago \citep{Riess_1998, Perlmutter_1999}.
The difficulties in reconciling the observed $\Lambda$ with expectations from fundamental physics \citep[see e.g.][]{Weinberg_1989} motivated scenarios in which the dark-energy equation of state evolves in time.
This may be supported by recent results from the DESI survey \citep{DESI_dr2}, although the suggested evolution seems difficult to reconcile with most models of dynamical dark energy.

These puzzles motivated the design of the current generation of survey facilities, the so-called ``Stage IV'' projects.
Among these, the DESI experiment \citep{desi_part_1,desi_part_2} is using a ground-based telescope to perform the largest spectroscopic survey to date.
Complementarily, the \textit{Euclid} space telescope stands out as the most comprehensive endeavour of this kind.
Euclid will collect imaging and spectroscopy in the visible and infrared bands over one third of the sky, to combine
galaxy clustering and weak gravitational lensing, as well as galaxy clusters and other probes,  
to attack the mysteries of the standard model with unprecedented precision and control of systematic errors \citep{Mellier_2025}.

Constraints on cosmological parameters from the large-scale distribution of objects in such huge surveys, are typically obtained through the computation of {\it summary statistics}, which are then compared to model predictions.
$N$-point correlation functions, or their equivalent in Fourier space, are used to quantify the observed deviations from homogeneity.
Two-point statistics, i.e., the correlation function or the power spectrum $P(\textbf{k})$, contain most cosmological information, yet it has become clear in recent years that the constraining power of the data can be significantly enhanced if higher-order functions are included in a joint inference \citep[see e.g.][]{Veropalumbo_2021}.
Accessing the full hierarchy of correlations not only yields tighter cosmological constraints, but in particular breaks degeneracies between cosmology and halo/galaxy formation parameters.

Alternatively, a complete field-level analysis of the data would in principle capture the full information at all levels of the hierarchy, bypassing the need for computing $N$-point summary statistics \citep[see e.g.][]{Leclercq_2021}.
This would also extract information about both the cosmological parameters and the initial conditions of our Universe.
A practical application to cosmological inference, however, has remained prohibitive so far.
This requires a forward model to generate realizations of density fields, which in general imply fixing the cosmology, in order to explore all possible realizations of initial conditions \citep[see e.g.][]{Jasche_2015,Lavaux_2019,Ata_2020}.

In this context, machine learning (ML) techniques have the potential to accelerate this process by associating a given realization of the Universe with the correct cosmological parameters.
Several studies over the past few years focused on cosmological inference using convolutional neural networks \citep[CNN -][]{LeCun_1989} applied to simulations \citep[see e.g.][]{Ravanbakhsh_2016,Gupta_2018,Ntampaka_2020,Ntampaka_2022,Villaescusa-Navarro_2022,Min_2024,Sharma_2024}, but also considering real data from weak lensing maps \citep[see e.g.][]{Jeffrey_2020,Jeffrey_2025}, or the galaxy distribution \citep[see e.g.][]{Lemos_2024}.
More recently, other ML techniques such as graph neural networks \citep[GNN - e.g.,][]{Battaglia_2018}, have also been applied to cosmological inference \citep[see e.g.][]{Makinen_2022,Villanueva_Domingo_2022,de_Santi_2023,Shao_2023,Balla_2024,Lee_2025}.
 
Cosmological inference using ML techniques, however, faces an intrinsic fundamental limitation: the standard machine-learning concept of ``training data'' does not make sense in cosmology.
Unlike the well-known cases of successful ML applications to language, imaging, or chemistry, cosmologists
cannot collect thousands of experimental measurements to train their ML algorithms.
Multiple realizations of the Universe under different cosmological parameters are simply not available.
As such, any cosmological inference programme based on ML must start from the problem of constructing the training samples, and is inevitably forced to resort to numerical simulations.
The requirements placed on the simulations in this respect, as to produce a reliable and effective training set, are twofold: (1) obviously, they have to be as realistic a reproduction of the real data as possible, to act as a robust surrogate of a true training set; (2) they must be produced in large numbers as to allow for a sufficiently precise training to be achieved. 

A program to build a set of simulations capable to fulfil these requirements and perform ML field-level inference from a galaxy survey catalogue has to face severe limitations. 
On one side, hydrodynamical simulations capable to follow the joint evolution of gravity and gas processes, directly generating galaxies, are computationally very expensive, thus limiting them in size and number. The best current example of a suite of small-volume hydrodynamical simulations designed to train ML tools is that produced by the \texttt{CAMELS} project \citep{Villaescusa-Navarro_2021}.
Purely gravitational $N$-body simulations have the advantage of allowing larger volumes to be explored, such as with the \texttt{Quijote} project \citep{Villaescusa-Navarro_2020}.
Yet, artificial galaxies have to be generated with sufficient realism within the dark matter haloes formed by the purely gravitational evolution. This is usually achieved through analytical methods, as halo occupation distribution models \citep[HOD --- see e.g.][]{Berlind_2002, Zheng_2005} or subhalo abundance matching \citep[SHAM --- see e.g.][]{Vale_2004, Conroy_2006}, which adds a further layer of uncertainty to the resulting mock samples.
Furthermore, even without hydrodynamics, the production of a sufficiently large training set of $N$-body simulations, with proper volume and resolution, still remains a computational demanding endeavour.

There are two alternatives to ameliorate this situation.
On one side, one can further reduce the simulation cost by approximating the gravity solver with perturbative methods, such as Lagrangian perturbation theory \citep[LPT --- see e.g.,][]{Monaco_2002}.
$N$-body simulations using the particle mesh method can also be accelerated with LPT-based time integration schemes \citep[see e.g.][]{Tassev_2013,Feng_2016, Leclercq_2020, Bartlett_2025, Rampf_2025}.
However, such codes usually work with fixed spatial grids and therefore do not have the same small-scale accuracy as adaptive methods, or other gravity solvers commonly used to produce high-resolution $N$-body simulations.
Efforts towards reproducing the density field of costly $N$-body simulations with ML methods are also ongoing \citep[see e.g.][]{He_2019,deOliveira_2020,Kaushal_2022,Jamieson_2023,Jamieson_2025}.

On the other hand, to improve the link between simulated and observational data, one could focus on cosmological objects and observables whose link to dark-matter haloes is simpler and more direct than for galaxies.
Clusters of galaxies, especially when selected in X rays, are a much closer realization of a dark matter halo and present several advantages in this respect.
Several applications of ML have been developed in the past to perform cosmological inference with galaxy clusters, focusing on other observables than the spatial clustering \citep[see e.g.][]{Qiu_2024, Cerardi_2025, Ntampaka_2025}.

Despite the difficulty of assembling large statistical samples going beyond historical ``eyeball" compilations in general not suited for robust cosmological studies, as, notably, the classic Abell catalogues \citep{Abell_1958,Abell_1989}, clusters of galaxies have thus their own advantages as probes of large-scale structure.
This includes, in particular, the ability to map the largest scales with limited samples and investment of telescope time.
X-ray selected clusters play a special role in these applications, given the more robust definition of these objects in X rays, and the direct link of X-ray emission to the cluster mass \citep[see][for a review]{Borgani_2001}.
In more recent years, a similarly robust approach is provided by the Sunyaev-Zel’dovich effect \citep[SZ -][]{Sunyaev_1972} produced by the inverse Compton scattering of Cosmic Microwave Background (CMB) photons on the hot ICM \citep[see, e.g., the review by][and references therein]{Clerc_2023}.
Large-area CMB surveys like ACT \citep{ACT_SZ_Clusters_2025}, SPT \citep{SPT_SZ_Clusters_2015}, and the Planck space mission \citep{PlanckXXVII_SZ_2025} delivered large catalogues of SZ clusters with a well-defined selection function.
This is one key feature, as for X-ray surveys, to be able to perform cosmological inference \citep[although see][for a combination of optical selection and weak lensing]{Fumagalli_2024}. 

Here, we shall focus specifically on X-ray selected catalogues, also in view of the interest in the application to the new all-sky catalogues by the eROSITA satellite mission \citep[e.g.][]{erosita_2024, Artis_2025, Ghirardini_2024}.
We also place ourselves under the least stringent observational setup of available X-ray surveys, in which the data to connect observed quantities to the cluster mass are limited.
This means assuming that X-ray fluxes are known (and thus luminosities), but that information as cluster ICM mean temperatures \citep[e.g.][]{Kravtsov_2012} or weak lensing data \citep[e.g.][]{Fumagalli_2024}, which could be used to estimate more reliable cluster masses, are not in general available.
This reproduces the most common situation in a generic all-sky X-ray survey of the past or current generation. 
Historically, the first reliable cosmological constraints from X-ray clusters were obtained using samples built from observations of the ROSAT satellite, either through analyses of their abundance and clustering \citep[as notably by the REFLEX project, ][]{boehringer04, collins00, guzzo09, schuecker02, schuecker03a, Balaguera_Antolinez_2011}, or from their evolution in deeper samples \citep[see, e.g.,][for a review]{Rosati_2002}.
These works still represent the main reference for clustering analyses of X-ray clusters \citep[although see e.g.][]{Marulli_2018}.
At the time of writing, in fact, cosmological results from the deeper eROSITA all-sky survey are limited to  cluster abundance \citep{Ghirardini_2024}.

In this paper, we present the first results from the {\it Machine Learning in Space} (MLS) project, a programme to explore cosmological inference using ML methods, when applied to catalogues of X-ray selected galaxy clusters.
While we are in parallel also exploring the use of graph neural networks for this scope, in this paper we focused on testing the ability of 3D convolutional neural networks to predict the values of the cosmological parameters from a combination of the abundance of clusters and field-level clustering information.
In particular, we train CNNs using large sets of synthetic cluster catalogues built from approximated, fast simulations.

As a benchmark, we perform a comparison with a neural network that instead takes in input the compressed information provided by the combination of number density (abundance) of clusters and their power spectrum, with the goal of assessing the potential of a CNN field-level approach to extract clustering information more efficiently than the binned power spectrum.
As discussed in the final section, the results presented here represent the first stage of what aims to become a fully-Bayesian pipeline to extract cosmological posteriors with a field-level approach based on ML, which will be developed in a forthcoming work.

The outline of this paper is as follows.
In \cref{sec:simu_dataset}, we present the generation of synthetic galaxy clusters mocks used to train the neural network models.
In \cref{sec:neural_net}, we present the architecture of the neural networks, describe the training procedure, and the hyperparameter optimization.
In \cref{sec:results}, we compare the performance of the different models in extracting cosmological parameters.
We conclude in \cref{sec:conclusion} with a discussion of future prospects and the limitations of the present analysis.

\section{Building the training samples}
\label{sec:simu_dataset}

\subsection{Dark matter haloes}
\label{subsec:dark_matter_haloes}

We constructed a suite of dark matter halo catalogues using \texttt{Pinocchio} \citep{Monaco_2002}, which uses 3rd order Lagrangian perturbation theory (3LPT) to generate approximate, but fast simulations of dark matter haloes.
The \texttt{Pinocchio} runs are initialized with the linear matter power spectrum at $z=0$, as computed by the \texttt{CAMB} Boltzmann solver \citep{Lewis_1999}.

We consider periodic cubic boxes of side length $1500\,h^{-1}\,\mathrm{Mpc}$ and a Cartesian grid with $750^3$ cells.
With this setup, we have a volume large enough to study the statistics of cluster-sized massive haloes, while being able to resolve them.
In fact, the smallest haloes formed in the simulations are made of $10$ particles, which for the cosmological model with the worst mass resolution that we consider ($\Omega_{\rm m}=0.5$) corresponds to a mass $\sim 10^{13}\,h^{-1}M_\odot$.
In this work, we focus on masses larger than $10^{14}\,h^{-1}M_\odot$.

The purpose of this large suite of mock catalogues is to train ML methods to extract the values of the cosmological parameters from a galaxy cluster catalogue.
For this, we sample the cosmological parameter space with a Sobol sequence \citep {Sobol_1967}.
This method has been used in the context of simulation-based inference using ML methods \citep[see e.g.][]{Bairagi_2025}, as well as to build cosmological emulators \citep[see e.g.][]{Kacprzak_2023,DeRose_2023,Chen_2025}.
One advantage of the Sobol sequence method over other commonly used sampling techniques, such as the Latin hypercube, is that it is always possible to further extend the number of points covering the parameter space.
Combined with the use of fast simulations, as in our case, this approach offers the flexibility to readily adjust the dataset, by progressively increasing its density according to the needs of the analysis.

We consider the five-dimensional parameter space of the standard $\Lambda$CDM model, represented by: (1) the total matter density parameter $\Omega_{\rm m}$; (2) the present-day root-mean-square linear matter fluctuations averaged over a sphere of radius $8\,h^{-1}\,\mathrm{Mpc}$ $\sigma_8$; (3) the Hubble parameter $h$; (4) the spectral index of the primordial power spectrum $n_{\rm s}$; and (5) the baryon density parameter $\Omega_{\rm b}$ .
We generate a Sobol sequence with $4096$ points covering the following parameter ranges:
\begin{align}
  0.1 & \leq \Omega_{\rm m} \leq 0.5, \\
  0.6 & \leq \sigma_{8} \leq  1, \\
  0.5 & \leq h \leq 0.9, \\
  0.8 & \leq n_{\rm s} \leq 1.2, \\
  0.03 & \leq \Omega_{\rm b} \leq 0.07.
\end{align}
For each of the $4096$ cosmologies in the Sobol sequence, we run a \texttt{Pinocchio} simulation using a different random seed for the initial conditions.

We also consider a fiducial cosmology (hereafter P$18$) with a \citet{Planck2018} compatible cosmology, where $\Omega_{\rm m}=0.3071$, $\sigma_8=0.8224$, $h=0.6803$, $n_{\rm s}=0.96641$, and $\Omega_{\rm b}=0.048446$.
For this cosmology, we produce $1\,000$ mocks with different initial conditions.
This set of mocks is used to estimate statistical errors for the different observational statistics considered in this work (see \cref{subsec:cosmo_probes}).

\subsection{Illuminating dark-matter haloes: the X-ray luminosity-mass relation}
\label{subsec:xray_lum_mass_relation}

\begin{figure}
    \centering
    \includegraphics[width=8cm]{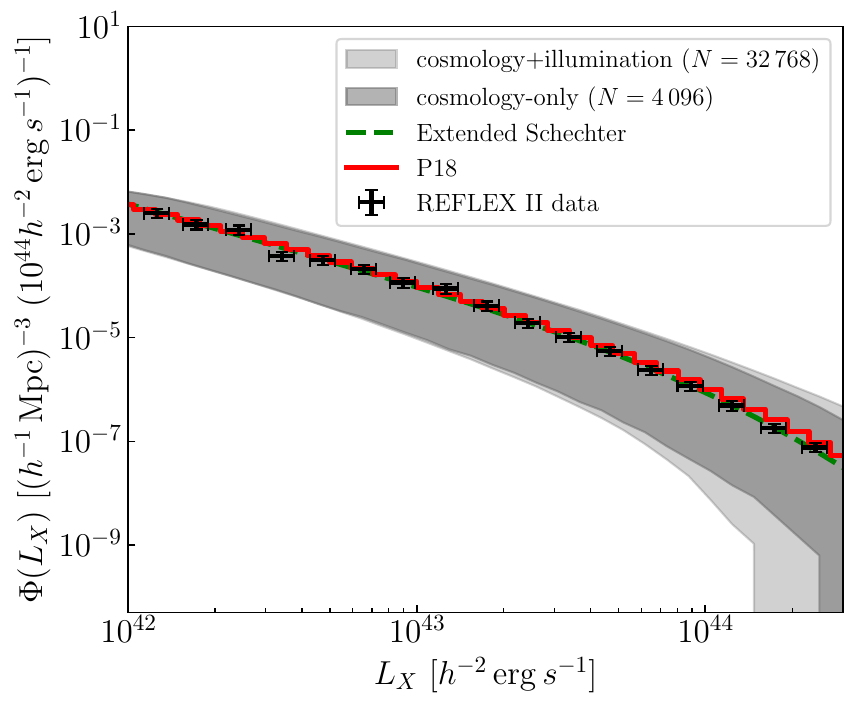}
    \caption{
    X-ray luminosity functions of the mock galaxy cluster catalogues, compared to the estimate from REFLEX-II \citep{boehringer14}. The dark shaded area correspond to $95.4\%$ interval of the region spanned by the $4096$ luminosity functions obtained from the mocks, when only the cosmological parameters are varied, i.e. fixing the  $M-L_{\rm X}$ relation parameters to the fiducial values. The lighter shaded area shows how this changes when the full set of $32768$ mocks, spanning the spread in the $M-L_{\rm X}$ relation parameters, is considered. X-ray luminosities are computed in the ROSAT $[0.1-2.4]$ keV band. The step line represents the reference cosmology model P18 and the dashed line corresponds to the fit using the extended Schechter function from \cref{eq:extendedSchechter}. Filled dots with error bars show the observed REFLEX-II luminosity function.
    }
    \label{fig:lx_mock_allcosmo_lumparams}
\end{figure}

We now turn to the construction of synthetic mocks mimicking X-ray selected clusters, based on the dark matter haloes presented in the previous section.
To reproduce in these the main observable, that is, the X-ray luminosity $L_{\rm X}$, we constructed a pipeline to “illuminate” the simulated dark matter haloes through a physically motivated recipe.
We base our procedure on the X-ray luminosity function from the REFLEX-II catalogue \citep{boehringer14, Balaguera_2012}, which is shown in \cref{fig:lx_mock_allcosmo_lumparams} and is well described by an extended Schechter function, i.e.
\begin{equation}
    \label{eq:extendedSchechter}
    \Phi(L_{\rm X})\,\mathrm{d}L_{\rm X} = n_0 \left(\frac{L_{\rm X}}{L_{\rm X}^\star}\right)^{-\alpha} e_q \left(-\frac{L_{\rm X}}{L_{\rm X}^\star}\right)\, \mathrm{d}\left(\frac{L_{\rm X}}{L_{\rm X}^\star}\right),
\end{equation}
where $n_0$ sets the normalization, $\alpha$ the low-luminosity slope, and $L_{\rm X}^\star$ the transition from power-law to the exponential cut-off.
The function $\rm e_q(x)$ is the $q$-exponential distribution \citep{Tsallis}, defined as
\begin{equation}
    \rm e_q(x) = 
    \begin{cases}
        e^x, & q = 1,\\
        [1 + x(1-q)]^{1/(1-q)}, & q \neq 1.
    \end{cases}
\end{equation}
The best-fit parameters for the REFLEX-II survey are $\alpha=1.54\pm0.06$, $L_{\rm X}^\star=(0.63\pm0.15)\times10^{44}\,h^{-2}\,\mathrm{erg}\,\mathrm{s}^{-1}$, $n_0=(4.08\pm0.82)\times10^{-6}\,(h^{-1}\,{\rm Mpc})^{-3}$, and $q=1.31\pm0.03$ \citep{Balaguera_2012}, where X-ray luminosities are measured in the ROSAT $[0.1-2.4]$ keV band.
The corresponding fit is shown as the dashed green line in \cref{fig:lx_mock_allcosmo_lumparams}.

In order to associate an X-ray luminosity to our dark matter haloes we follow the procedure from \citet{Balaguera_2012}, which uses an empirical $M-L_{\rm X}$ scaling relation with a log-normal scatter.
The mean of the scaling relation is parametrized as a quadratic function, given by
\begin{equation}
    \label{eq:BA_quadratic}
    \bar{\ell} = a_{\rm X} + b_{\rm X}\,m + c_{\rm X}\,m^2,
\end{equation}
where $\bar{\ell} = \log_{10}(\overline{L}_{\rm X} / 10^{44}\,h^{-2}\,\mathrm{erg}\,\mathrm{s}^{-1})$, with $\bar{L}_{\rm X}$ the mean of the log-normal relation, and $m = \log_{10}(M/10^{14}\,h^{-1}\,M_\odot)$.
Here, $a_{\rm X}$, $b_{\rm X}$, $c_{\rm X}$ are free parameters, assumed to be redshift independent.
This assumption is valid for cluster samples such as REFLEX or REFLEX II, where the majority of the objects that can be used to study the large-scale structure are at $z<0.2$.
Redshift evolution, however, would need to be included for deeper cluster samples as those of the recent eROSITA survey \citep{erosita_2024}, which is simple to do with the current model.

The scatter of the log-normal scaling relation $\sigma_{\rm tot}$ is made of two contributions that can be added in quadrature: an intrinsic scatter $\sigma_{\ln L|M}$ and an observational uncertainty of $\sigma_{\rm obs}$.
The observational uncertainty for the REFLEX II survey has been estimated by \citet{Balaguera_2012} to be $\sigma_{\rm obs}=0.2$.

In \citet{Balaguera_2012} they fitted the mean of the $M-L_{\rm X}$ scaling relation, as described in \eqref{eq:BA_quadratic}, to the REFLEX II data using $N$-body simulations assuming fixed values for the cosmological parameters.
They obtained $a_X=-1.36\pm0.03$, $b_X=1.88\pm0.05$, $c_X=-0.29\pm0.04$.
On top of this, they considered an intrinsic scatter of $\sigma_{\ln L|M} = 0.26$, as estimated by \citet{Stanek2010} from the \citep{Reiprich2006} subset of REFLEX luminous clusters.
In the following we will refer to this parameter values as the fiducial $M-L_{\rm X}$ scaling relation parameters.

In \cref{fig:lx_mock_allcosmo_lumparams}, we show the X-ray luminosity function measured from one realization of our P$18$ cosmology, using the fiducial values of the $M-L_{\rm X}$ relation.
We use a single snapshot at $z=0.1$, which is close to the median redshift of the RELFEX II sample.
We see a good agreement with both the data from REFLEX II and the extended Schechter function.
We also apply the fiducial illumination procedure to the $4096$ mocks from the Sobol sequence varying the five $\Lambda$CDM parameters.
The $95.4\%$ confidence intervals of the resulting distribution of X-ray luminosity functions is displayed \cref{fig:lx_mock_allcosmo_lumparams}.
This represents the impact of the cosmological parameters on the X-ray luminosity function, at fixed $M-L_{\rm X}$ scaling relation parameters.

For our inference procedure we do not use the fiducial values of the $M-L_{\rm X}$ parameters.
Indeed, we want to be able to marginalize over the uncertainty of the $M-L_{\rm X}$ parameters.
Additionally, we want to be able to self-calibrate our own $M-L_{\rm X}$ scaling relation, that is consistent with our dark matter halo catalogues, and is independent from the assumption of a fiducial cosmology. 
To do that, the parameters $(a_{\rm X}, b_{\rm X}, c_{\rm X}, \sigma_{\rm tot})$ have been sampled following a Sobol sequence, with bounds set to six times the scatter estimated by \citet{Balaguera_2012} around their fiducial values.
This allows us to efficiently explore the parameter space while remaining consistent with the REFLEX-II data.
More specifically, for each of the $4096$ cosmological models, we generate an independent Sobol sequence with $8$ samples for the four $M-L_{\rm X}$ relation parameters $(a_{\rm X}, b_{\rm X}, c_{\rm X}, \sigma_{\rm tot})$.
We therefore obtain a total of $32\,768$ mocks sampling the $(5+4)$-dimensional parameter space.

We represent the X-ray luminosity function from the  $32\,768$ X-ray cluster mocks in \cref{fig:lx_mock_allcosmo_lumparams}.
More specifically, we give the $95.4\%$ confidence intervals of the distribution of possible X-ray luminosity functions.
This shows the impact of varying the cosmological parameters and the $M-L_{\rm X}$ relation parameters simultaneously.

\subsection{Methodology: extracting cosmological information from cluster catalogues}
\label{subsec:cosmo_probes}

As a starting point, we focus on a single simulation snapshot at $z=0.1$, with a fixed comoving volume of $\left(1\,500\,h^{-1}\,\mathrm{Mpc}\right)^3$.
We select clusters with a luminosity $L_{\rm X} \geq 3 \times 10^{43}\,h^{-2}\,\mathrm{erg}\,\mathrm{s}^{-1}$ in the ROSAT $[0.1,2.4]$ keV band, which roughly corresponds to the sensitivity of the REFLEX survey at $z=0.1$.
In the fiducial cosmology P$18$, and with the fiducial $M-L_{\rm X}$ relation parameters, we obtain a sample of $\sim 14\,500$ clusters, again consistent with the observed REFLEX X-ray luminosity function \citep{boehringer02, boehringer14}.
As previously described, we consider different statistics in order to extract cosmological information from the cluster catalogue, using both the abundance and the clustering of clusters.
It is important to stress that also when considering standard summary statistics alone, the analysis is also based on a ML approach, training a multi-layer neural network, not the classic Bayesian likelihood procedure.
We restrict all clustering analysis to scales $k\lesssim 0.13 \,h\,\mathrm{Mpc}^{-1}$.
At such a scale, the power spectrum of the cluster sample in the fiducial mock is close but above the shot noise level (see \cref{fig:power_spectrum_sobol}).  
We now provide a technical description of each observational statistic considered.

\begin{itemize}

    \item \textbf{Cluster abundance (i.e., number density)}.
    We compute the number of clusters within each comoving simulation cube with an X-ray luminosity larger than a given threshold $L_{\rm X} \geq L_{\rm X,th}$. 
    Observationally, this would coincide to using the integrated X-ray luminosity function of a sample of clusters.
    We consider $10$ thresholds between $L_{\rm X,th}^1=3 \times 10^{43}\,h^{-2}\,\mathrm{erg}\,\mathrm{s}^{-1}$ and $L_{\rm X,th}^{10}=3 \times 10^{44}\,h^{-2}\,\mathrm{erg}\,\mathrm{s}^{-1}$, with a constant logarithmic spacing of $\Delta \log_{10}{L_{\rm X,th}}=0.1$.
    In \cref{fig:number_counts_sobol}, we show the distribution of the cluster abundance measurements from the whole set of $32\,768$ mocks, varying the    cosmological and $M-L_{\rm X}$ relation parameters.
    As a reference, the measurement for the fiducial P$18$ mock is also given.

    \begin{figure}
    \centering
    \includegraphics[width=8cm]{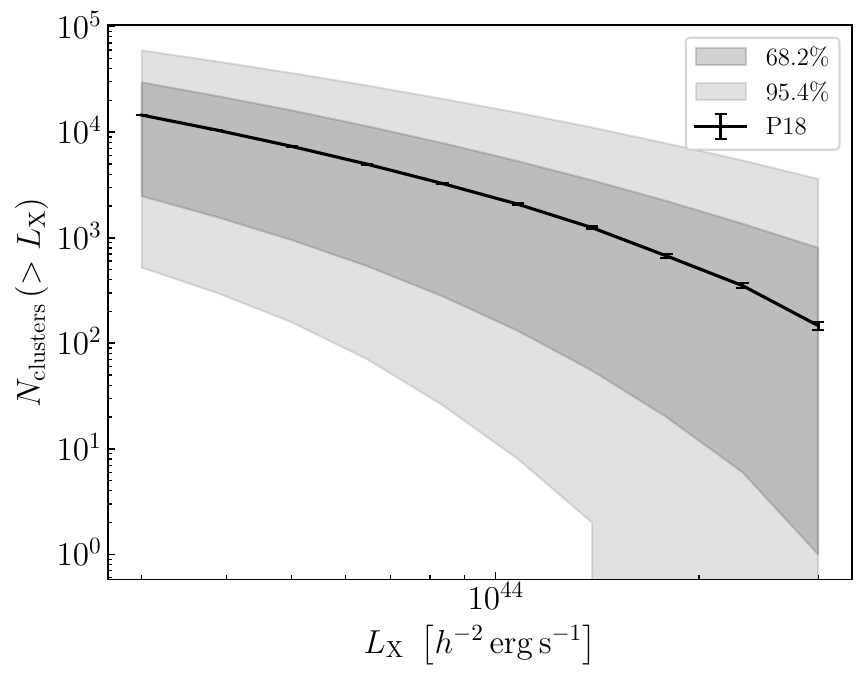}
    \caption{
        Number of clusters as a function of the X-ray luminosity in cumulative bins measured.
        The solid line gives the measurements for the first realization of the P$18$ cosmology with the fiducial $M-L_{\rm X}$ relation parameters, with the associated statistical errors estimated from the $1\,000$ realizations (not always visible).
        The dark and light shaded areas delimit the $68.2\%$ and $95.4\%$ confidence intervals of the distribution of measurements from the whole set of $32\,768$ mocks, varying both cosmological and $M-L_{\rm X}$ relation parameters.
        }
    \label{fig:number_counts_sobol}
    \end{figure}

    \item \textbf{Power spectrum of the cluster distribution}.
    To compute the power spectrum from the cluster catalogue, we need to derive an overdensity field from the distribution of discrete objects, which is then fed to a Fast Fourier Transform.
    To this end, we use a $64^3$ grid with a piecewise cubic spline (PCS) assignment scheme.
    We use the \texttt{pypower}\footnote{\url{https://github.com/cosmodesi/pypower}} library to measure the power spectrum, with second order interlacing \citep{Sefusatti_2016}.
    We sample $P(k)$ with $92$ equidistant bins in the wavenumber $k$, ranging from the fundamental mode of the box $k_{\rm f}\simeq 4\times 10^{-3}\,h\,\mathrm{Mpc}^{-1}$ to the Nyquist frequency of the grid $k_{\rm Nyq.}\simeq1.3\times 10^{-1}\,h\,\mathrm{Mpc}^{-1}$.
    We use a constant bin width of $\Delta k=0.33\,k_{\rm f}$.
    The choice of the bin width is explained in \ref{app:ps_binning}.
    We do not subtract the shot noise from our power spectrum measurements, since we will always use it in combination with the abundance of clusters.
    The power spectrum is computed for the full sample of clusters with luminosity $L_{\rm X} \geq 3 \times 10^{43}\,h^{-2}\,\mathrm{erg}\,\mathrm{s}^{-1}$.
    In \cref{fig:power_spectrum_sobol}, we present the distribution of power spectra measured from our mocks.
    As a reference, we also show the fiducial P$18$ case, with its associated shot noise level.

    \begin{figure}
    \centering
    \includegraphics[width=8cm]{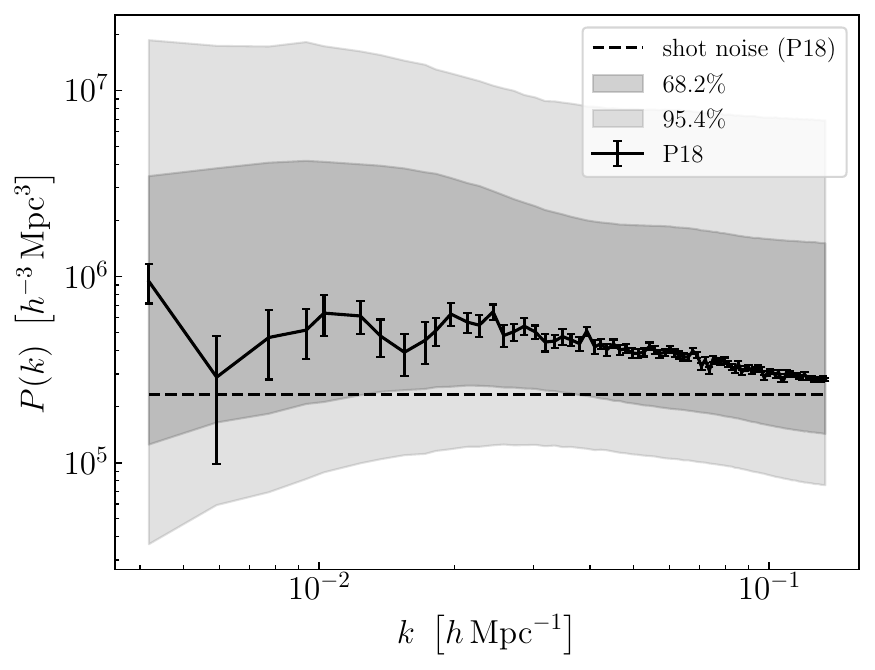}
    \caption{
        Power spectrum measurements from the mocks.
        The solid line gives the power spectrum for the first realization of the P$18$ cosmology with the fiducial $M-L_{\rm X}$ relation parameters, with the associated statistical errors estimated from the $1\,000$ realizations (not always visible).
        The dashed line gives as a reference the shot noise level of this fiducial case.
        The dark and light shaded areas delimit the $68.2\%$ and $95.4\%$ confidence intervals of the distribution of power spectra from the whole set of $32\,768$ mocks, varying both cosmological and $M-L_{\rm X}$ relation parameters.
    }
    \label{fig:power_spectrum_sobol}
    \end{figure}

    \item \textbf{Field-level analysis with CNN.}
    Field-level methods as CNNs need a continuous (pixelized) field for their application.
    Thus, we compute an overdensity field by interpolating the cluster positions into a Cartesian grid with $64^3$ cells using a PCS assignment scheme.
    More specifically, the overdensity field is defined as
    \begin{equation}
        \delta\left(\mathbf{x}\right) = \frac{\rho\left(\mathbf{x}\right)}{\bar{\rho}} - 1,
    \end{equation}
    where $\rho\left(\mathbf{x}\right)$ is the density at the comoving position $\mathbf{x}$ and $\bar{\rho}$ is the mean density over the whole simulation volume.
    We are using the same grid as for the power spectrum, ensuring in this way that the CNN and $P(k)$ analyses are probing the same range of scales.
    We consider two cuts in luminosity: the full sample with $L_{\rm X} \geq 3 \times 10^{43}\,h^{-2}\,\mathrm{erg}\,\mathrm{s}^{-1}$ and a high luminosity sample with $L_{\rm X} \geq 1.08 \times 10^{44}\,h^{-2}\,\mathrm{erg}\,\mathrm{s}^{-1}$.
    Additionally, we compute a weighted overdensity field, where each individual cluster is weighted by $w_{\rm X}^i=L_{\rm X}^{i}/\bar{L}_{\rm X}$, where $L_{\rm X}^{i}$ is the luminosity of an individual cluster and $\bar{L}_{\rm X}$ is the mean luminosity computed over the whole catalogue.
    This is the same weighting scheme as used in \citet{Balaguera-Antolinez_2014} in the context of a marked power spectrum analysis.

\end{itemize}

We stress that in this analysis we neglect both redshift-space distortions \citep[RSD;][]{Kaiser_1987} and the Alcock–Paczynski effect \citep{Alcock_1979}.
Thus, clusters are placed at their real-space comoving positions, which are known \textit{a priori} from the simulation, not considering that actual observed data would be affected by peculiar velocities (and thus will be in redshift space) and will only provide us with angles and redshifts, requiring the assumption of a fiducial cosmology to convert to actual distances.
Incorporating these observational effects will be necessary before applying the method to real data.
Nevertheless, both can be straightforwardly implemented within the current pipeline, the main adjustment being the inclusion of the power spectrum quadrupole to ensure a fair comparison.
In any case, we do not expect these effects to substantially alter the information content, at least for the class of tracers considered in this analysis, although they can bias the inferred cosmological parameters if not properly accounted for \citep[see e.g.][]{Fumagalli_2025}.

\section{Neural networks}
\label{sec:neural_net}

\subsection{Architecture}
\label{subsec:architecture}

\begin{figure*}
    \centering
    \includegraphics[width=0.9\linewidth]{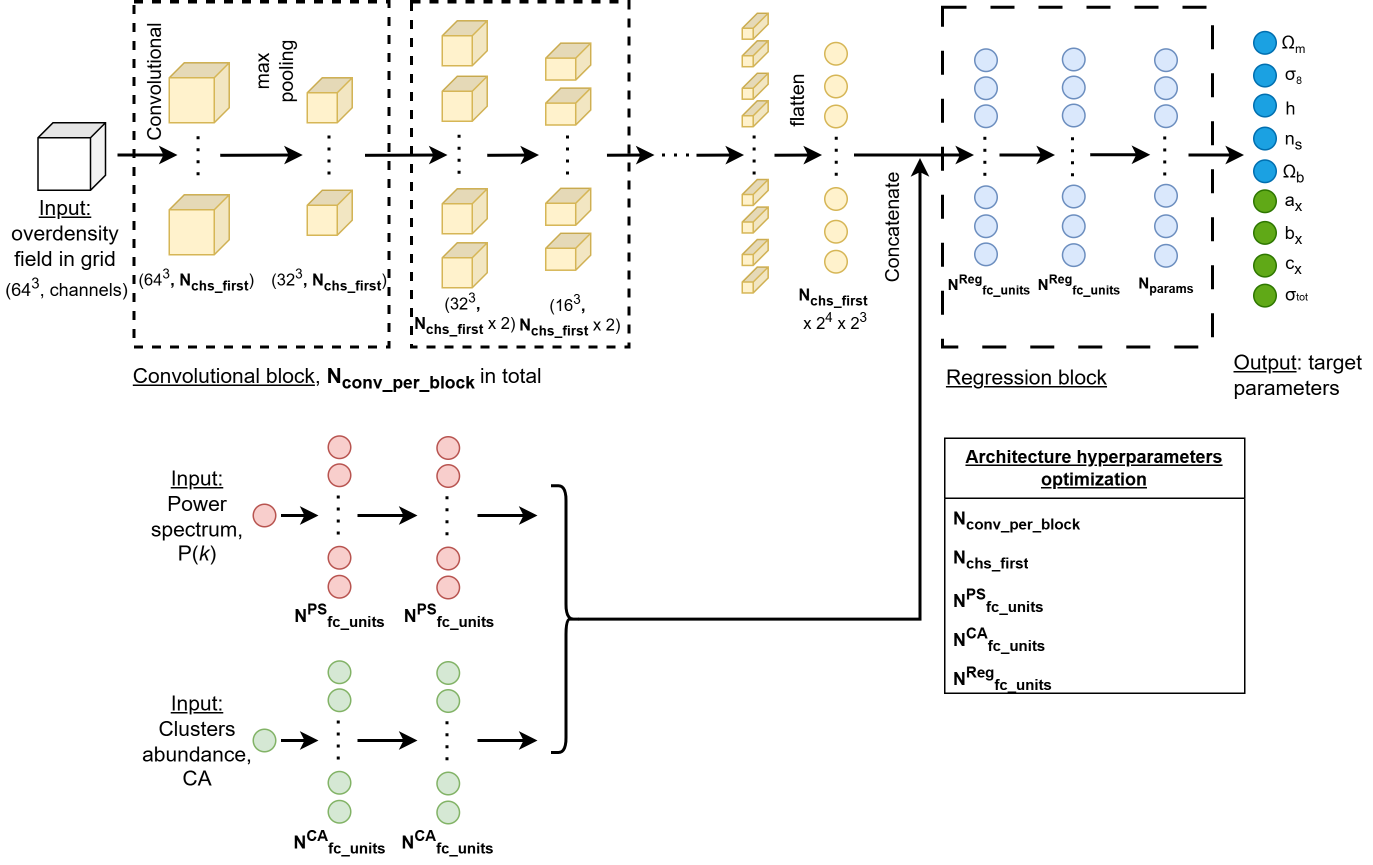}
    \caption{
        General architecture of the deep neural networks used for parameter estimation.
        A set of feature extraction networks extracts in parallel features for each of the considered observational statistics.
        The CNN used to extract features at the field level takes as input a tensor of shape ($64^3$, channels).
        The CNN module is an encoder that consists of five consecutive convolution blocks with $N_{\rm conv\_per\_block}$ convolution layers each and $N_{\rm chs\_first}$ output channels for the first block.
        Each subsequent block multiplies the number of output channels by two with respect to the previous one.
        For cluster abundance and power spectrum, the extraction network consists of two fully connected (FC) layers with $N_{\rm fc\_units}$ neurons per layer.
        The extracted features are concatenated into a one-dimensional array that feeds a final regression block that outputs the target parameters.
        Different combinations of input observational statistics can be chosen to produce different inference models.
        The output of the network is a vector of size $N_{\rm params}$, corresponding to the $5+4$ cosmological and $M-L_{\rm X}$ relation parameters.
        A more detailed description of the architecture is given in \cref{subsec:architecture}.
        }
    \label{fig:NNarchitecture}
\end{figure*}

We build neural networks whose aim is to predict the cosmological and $M-L_{\rm X}$ relation parameters, based on different input observational statistics.
These networks are divided in several components.
\Cref{fig:NNarchitecture} schematically depicts the general architecture.
First, a set of feature extraction networks derive features in parallel from the different observational statistics that we consider, that is, the cluster abundance, the power spectrum, or the field-level analysis with the CNN.
We use several of these networks in parallel depending on the particular combination of statistics that we consider.
Then, the output of the different feature extraction networks are concatenated into a one-dimensional array, which is passed to a final regression network.
The final regression network makes the link between the combined extracted features and the target cosmological parameters.
We note that the general structure is very similar to the one used in \citet{Min_2024}, although we are considering different statistics in this work.
The networks have been implemented and trained using \texttt{PyTorch}~\citep{pytorch} and our code is made publicly available\footnote{\url{https://github.com/ICSC-Spoke3/MLS-CNN}}.

\textbf{Cluster abundance and power spectrum analysis.} In order to extract features from the number density of clusters and their power spectrum, we use the same architecture based on a dense fully connected (FC) neural network.
This includes $2$ hidden linear layers with $N_{\rm fc\_units}$ neurons per layer.
Each hidden layer is followed by a batch normalization layer \citep{ioffe2015batch}, a \texttt{ReLU} activation layer, and a dropout layer with rate $p_{\rm drop}$.
The input for the feature extraction networks is a one-dimensional array containing either the number of clusters in X-ray luminosity bins or their power spectrum in wavenumber bins.

\textbf{Field-level analysis with CNN.} In order to extract features from the overdensity field we use a 3D convolutional neural network.
The CNN chains multiple convolutions which can be divided into different blocks.
Each convolutional block is made of $N_{\rm conv\_per\_block}$ convolutional layers with a kernel of size $3\times3\times3$, a stride of $1$, and a zero padding of $1$.
This kernel configuration preserves the spatial shape of the input tensor.
Each convolutional layer is followed by a batch normalization layer and a \texttt{ReLU} activation layer.
After the last convolution in a block, a max pooling layer with a kernel of size $2\times2\times2$, a stride of $2$, and padding of $0$, is used to divide the size of the field by a factor of two in each spatial dimension.
We use $5$ convolutional blocks, so that the input spatial dimension of shape $64^3$ is transformed into a field of shape $2^3$.
The first convolutional block uses $N_{\rm chs\_first}$ output channels in each of the $N_{\rm conv\_per\_block}$ convolutions.
In each subsequent block, the number of output channels doubles that of the previous block.
The final number of channels is therefore $N_{\rm chs\_first}\times2^4$.
The output of the last block is flattened into a one-dimensional array of size $N_{\rm chs\_first}\times2^4\times2^3$.
The CNN can take as input multiple three-dimensional overdensity fields, combined as separate input channels.
This enables, for instance, the combination of overdensity fields computed for different cuts in X-ray luminosity.

\textbf{Combination of statistics and final regression.}
The features extracted from the abundance of clusters, and either the power spectrum or the field-level analysis with the CNN, are then concatenated into a single one-dimensional array and passed to a final dense fully connected network.
Its architecture is almost identical as the one used to extract features from the abundance of clusters and power spectrum previously described.
The only addition is a final linear layer, with no activation function, batch normalization or dropout layer after it.
The output of this layer is the final output of the full network, which predicts the values of the cosmological and $M-L_{\rm X}$ relation parameters.

\subsection{Training}
\label{subsec:training}

We split our data into three different sets: a training dataset with $28\,672$ samples ($87.5\%$), a validation dataset with $2048$ samples ($6.25\%$), and a test dataset with $2048$ samples ($6.25\%$).
The validation dataset is used to optimize the hyperparameters of the networks, as described in \cref{subsec:hp_optim}, as well as to prevent overfitting with an early stopping mechanism.
The test dataset is reserved to evaluate the final performance of each model (see \cref{subsec:performance_test_set}).

We standardize the training dataset before passing it to the networks.
Each cosmological parameter is individually rescaled to a zero-mean and unit variance distribution.
For the cluster abundance and power spectrum, we work with $\log_{10}{\left(1+N_{\rm clusters}\right)}$ and $\log_{10}{\left(P\right)}$, respectively.
This is done to reduce the dynamical range of the training samples.
Then, we subtract the mean and divide by the standard deviation computed over all training samples and luminosity or wavenumber bins.
For the CNN inputs, we subtract to each pixel the mean and divide by the standard deviation computed over all training samples and pixels.

In order to train the networks, we consider a mean squared error loss function, which can be written as
\begin{equation}
    \label{eq:loss_mse}
    \mathcal{L}_{\rm MSE} = \frac{1}{N_{\rm batch}}\sum_{i=1}^{N_{\rm batch}}\frac{1}{N_{\rm params}} \sum_{j=1}^{N_{\rm params}} \left(x_{i,j}-y_{i,j}\right)^2,
\end{equation}
where $x_{i,j}$ is the network prediction for parameter $j$ and sample $i$, $y_{i,j}$ is the target parameter $j$ of sample $i$, $N_{\rm batch}$ is the batch size, and $N_{\rm params}$ is the number of target parameters.
We minimize the loss function using the \texttt{AdamW} optimizer \citep{Kingma_2014,Loshchilov_2017}.
The learning rate is controlled with the \texttt{ReduceLROnPlateau} scheduler that reduces the learning rate by a constant factor whenever the validation loss has not improved for more than $10$ successive epochs.
We let the model train for a total maximum number of $1000$ epochs.
On top of this, we use an early stopping criterion that stops the model training whenever the validation loss has not improved for $20$ successive epochs.
This mechanism is meant to avoid overfitting the training set.
In practice, most model training runs are stopped by the early stopping criterion and never reach the maximum number of epochs. In most cases, the final total number of epochs is of $\sim 100-200$.
Once the training is finished, we save the state of the model at the epoch with the minimum validation loss, which might be different from the last epoch.
The networks are trained using NVIDIA A$100$ and H$100$ GPUs with $80$GB of memory. On a single of such GPUs, the training of one CNN model takes $\sim 1-3$ hours.

\subsection{Hyperparameter optimization}
\label{subsec:hp_optim}

\begin{table}
    \centering
    \begin{tabular}{c|c|c|c|c}
        Parameter & Min. value & Max. value & Step & Sampling \\
        \hline \hline
         $\gamma_0$ & $10^{-5}$ & $10^{-2}$ & - & log \\
         $\gamma_{\rm r}$ & $0.01$ & $0.9$ & - & log \\
         $\beta_1$ & $0.85$ & $0.999$ & - & linear \\
         $\beta_1$ & $0.99$ & $0.9999$ & - & linear \\
         $\epsilon$ & $10^{-8}$ & $10^{-4}$ & - & log \\
         $\lambda_{\rm w}$ & $10^{-6}$ & $10^{-2}$ & - & log \\
         $\log_{2}{N_{\rm batch}}$ & 6 & 9 & - & linear \\
         $N_{\rm chs\_first}$ & 1 & 10 & 1 & log \\
         $N_{\rm conv\_per\_block}$ & 1 & 2 & 1 & log \\
         $N^{\rm PS}_{\rm fc\_units}$ & 100 & 1000 & 100 & linear \\
         $N^{\rm CA}_{\rm fc\_units}$ & 100 & 1000 & 100 & linear \\
         $N^{\rm Reg}_{\rm fc\_units}$ & 100 & 1000 & 100 & linear \\
         $p_{\rm drop}$ & $0$ & $0.1$ & - & linear \\
    \end{tabular}
    \caption{
        Summary of the optimized hyperparameters. 
        The definition of each hyperparameter is given in \cref{subsec:architecture,subsec:hp_optim}.
        The minimum and maximum values of the search space, as well as the step size for the integer hyperparameters, are provided.
        We also indicate whether a linear of log sampling is used.
    }
    \label{tab:hp_opt}
\end{table}

We optimize the hyperparameters of the network with the \texttt{Optuna} library \citep{optuna_2019}.
We use the validation loss at the last epoch of a given trial as the metric to be minimized.
The hyperparameter space is sampled with the Tree-structured Parzen Estimator \citep[TPE --- see e.g.][]{Bergstra_2011, Bergstra_2013, Watanabe2023}.
The TPE sampler suggests hyperparameter combinations that are expected to produce an improved validation loss based on the knowledge of the previously completed trials.
We use the median pruner to kill unpromising trials, which stops trials whose validation loss at a given epoch is worse than the median of previously completed trials at the same epoch.
For the first $10$ trials, the sampling of the hyperparameters space is carried out randomly and no pruning is used.
Afterwards, new configurations are suggested by the TPE sampler and the median pruner is activated.
We run the optimization procedure for a total of $100$ trials.

In terms of learning rate, we optimize the value of the initial learning rate $\gamma_0$ and the reduction factor $\gamma_{\rm r}$ of the \texttt{ReduceLROnPlateau} scheduler.
For \texttt{AdamW}, we optimize the values of the $\beta_1$ and $\beta_2$ coefficients, the $\epsilon$ parameter, and the weight decay coefficient $\lambda_{\rm w}$.
We refer the reader to \citet{Loshchilov_2017} for the exact definitions of these parameters.
We also optimize the batch size $N_{\rm batch}$ used for the loss function updates.

For the CNN, we optimize the number of output channels of the first convolutional layer $N_{\rm chs\_first}$ and the number of convolutional layers per block $N_{\rm conv\_per\_block}$.
Since larger values of these hyperparameters result in networks with more parameters, which are more expensive to train, we impose a prior that favours lower values.
In practice, we sample these hyperparameters from the log domain, which makes the sampler suggest smaller values more often than larger values.
We refer the reader to the \texttt{optuna} documentation\footnote{\href{https://optuna.readthedocs.io/en/stable/reference/generated/optuna.trial.Trial.html\#optuna.trial.Trial.suggest_int}{https://optuna.readthedocs.io}} for the details of the procedure used to sample an integer parameter in the log domain.

For the fully connected networks, we optimize the number of units per hidden layer for the power spectrum $N^{\rm PS}_{\rm fc\_units}$, the cluster abundance $N^{\rm CA}_{\rm fc\_units}$, and the final regression network $N^{\rm Reg}_{\rm fc\_units}$.
We also optimize the value for the dropout rate $p_{\rm drop}$ common to all fully connected networks.
The optimized hyperparameters, with their search space range, are summarized in \cref{tab:hp_opt}.

\section{Results}
\label{sec:results}

\begin{table}
    \centering
    \begin{tabular}{c|c}
        Model & Number of trainable parameters \\
        \hline \hline
         CA+PS & $3\ 284\ 909$ \\
         CNN & $758\ 499$ \\
         CA+CNN & $1\ 569\ 699$ \\
         CA+CNN+PS & $2\, 552\, 799$ \\
         CA+$\mathrm{CNN}_{w_{\rm X}}$ & $3\ 804\ 799$ \\
         CA+$\mathrm{CNN}_{\rm 2x}$ & $3\ 510\ 169$ \\
    \end{tabular}
    \caption{
        Number of trainable parameters for each neural network model considered once the hyperparameter tuning is completed.
    }
    \label{tab:model_free_params}
\end{table}

We trained different neural network models corresponding to different combinations of observational statistics. The baseline case is the combination of cluster abundance (or equivalently, mean density) and the power spectrum of their spatial distribution.
We then explored several CNN-based scenarios, combining the field-level information with that provided by the abundance of clusters.
The driving idea is always that of testing how much more information the CNN is capable of extracting with respect to the simple power spectrum:
\begin{itemize}
    \item CA+PS: classic summary-statistics based analysis combining the cluster abundance, that is, the number density of clusters in X-ray luminosity bins (in practice, the X-ray luminosity function), with the power spectrum of the full sample with $L_{\rm X} \geq 3 \times 10^{43}\,h^{-2}\,\mathrm{erg}\,\mathrm{s}^{-1}$.
    \item CNN: analysis of the overdensity field based on the CNN for the full sample with $L_{\rm X} \geq 3 \times 10^{43}\,h^{-2}\,\mathrm{erg}\,\mathrm{s}^{-1}$.
    \item CA+CNN: combination of the cluster abundance and the CNN analysis.
    \item CA+CNN+PS\footnote{This model is expensive both in terms of memory and computing time. We therefore simplify the hyperparameter optimization procedure with respect to the other models. First, we fix $\log_2N_{\rm batch}=6$. In all our tests we have found that the batch size does not play a significant role for the final accuracy of the model. Second, we fix $N_{\rm conv\_per\_block}=1$, which is the optimum value found for all the other CNN-based models considered.}: combining cluster abundance, CNN-based analysis, and power spectrum.
    \item CA+$\mathrm{CNN}_{w_{\rm X}}$: same as CA+CNN, but now weighting the overdensity field by X-ray luminosity, as described in \cref{subsec:cosmo_probes}.
    \item CA+$\mathrm{CNN}_{\rm 2x}$: combining the cluster abundance with two unweighted overdensity fields obtained from the usual $L_{\rm X} \geq 3 \times 10^{43}\,h^{-2}\,\mathrm{erg}\,\mathrm{s}^{-1}$ sample and a high-luminosity set with $L_{\rm X} \geq 1.08 \times 10^{44}\,h^{-2}\,\mathrm{erg}\,\mathrm{s}^{-1}$. The two fields are combined as different input channels to the CNN.
\end{itemize}

As described in \cref{sec:neural_net}, the training and hyperparameter tuning is performed independently for each of the models considered.
The number of trainable parameters corresponding to the best hyperparameter configuration obtained after the tuning procedure is given in \cref{tab:model_free_params}.

\subsection{Metrics}
\label{subsec:metrics}

We use different metrics to quantify the performance on each trained model, when applied to the test set mocks.
We compute these metrics individually for each of the output parameters of the networks, that is, the target cosmological and the $M-L_{\rm X}$ scaling relation parameters.
The main quantity that we use to compare how well different models can extract cosmological parameters is the mean absolute relative error, defined as
\begin{equation}
    \label{eq:mean_abs_rel_err}
    \epsilon = \frac{1}{N_{\rm test}}\sum_{i=1}^{N_{\rm test}}\left|\frac{y_{\rm pred}^{i}-y_{\rm true}^{i}}{y_{\rm true}^{i}}\right|,
\end{equation}
where $N_{\rm test}$ is the number of samples in the test set, $y_{\rm pred}$ is the prediction of the network for a given target parameter, and $y_{\rm data}$ is the true value of the target parameter.
We also consider the normalized mean bias, defined as
\begin{equation}
    \label{eq:mean_bias}
    \mathrm{b} = \frac{1}{N_{\rm test}} \sum_{i=1}^{N_{\rm test}} \frac{y_{\rm pred}^{i}-y_{\rm true}^{i}}{\bar{y}_{\rm true}},
\end{equation}
where $\bar{y}_{\rm true}$ is the mean computed over the test set.
This quantity allows us to estimate the average bias across the considered parameter space.
Finally, we compute the $R^2$ score or coefficient of determination, which quantifies the quality of the regression performed by the neural networks.
It is defined as
\begin{equation}
    \label{eq:r2score}
    R^2 = 1 - \frac{\sum_{i=1}^{N_{\rm test}} (y^i_{\rm true} - y^i_{\rm pred})^2}{\sum_{i=1}^{N_{\rm test}} (y^i_{\rm true} - \bar{y}_{\rm true})^2}.
\end{equation}
A value of one indicates a perfect regression.

\subsection{Performance on the test set}
\label{subsec:performance_test_set}

\begin{figure*}
    \centering
    \includegraphics[width=\linewidth]{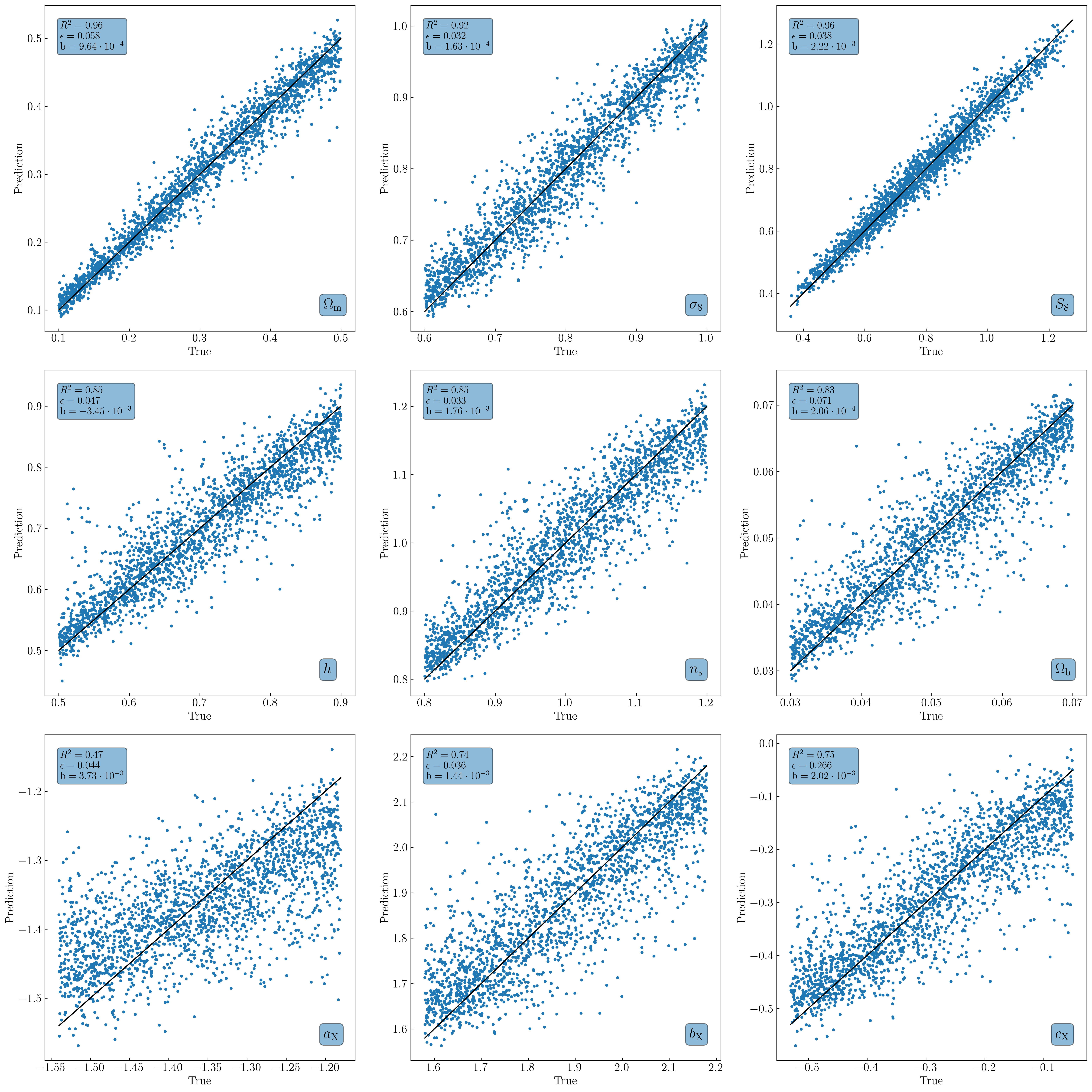}
    \caption{
        Test set predictions of the neural network compared to the true target parameters, for the case of the CA+CNN model.
        Each panel focuses on a different parameter.
        We omit the parameter $\sigma_{\rm tot}$ since it is poorly constrained.
        We have added the derived parameter $S_8$.
    }
    \label{fig:test_set_eval_mosaic_cnn}
\end{figure*}

\Cref{fig:test_set_eval_mosaic_cnn} gives the predicted value of each cosmological parameter compared to the true values for the CA+CNN model, for the mocks from the test set.
Each panel focuses on a different cosmological or $M-L_{\rm X}$ scaling relation parameter and presents the associated performance metrics.
We do not show the results for the parameter $\sigma_{\rm tot}$, since we find it unconstrained in all the models considered.
We will therefore ignore it for the rest of the paper.
Nevertheless, $\sigma_{\rm tot}$ is still accounted for during the procedure, which means that the results obtained for the other parameters are marginalized over it.
Additionally, we compute the parameter $S_{8}=\sigma_{8}\sqrt{\Omega_{\rm m}/0.3}$ as a derived quantity.

\begin{figure}
    \centering
    \includegraphics[width=8cm]{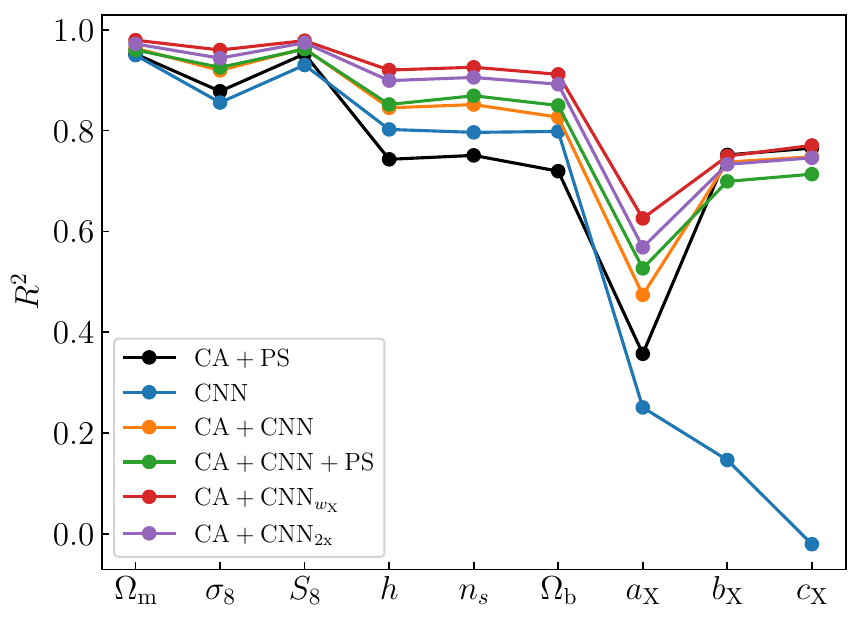}
    \caption{
        $R^2$ score evaluated on the test set for each target parameter.
        We have also included the derived parameter $S_8$.
        Each colour represents a different combination of statistics.
        We exclude $\sigma_{\rm tot}$, since it is unconstrained in all the cases considered.
    }
    \label{fig:r2score_cnn_vs_ps_test_set}
\end{figure}

For all parameters considered, the normalized mean bias is smaller than $10^{-2}$ with some cases smaller than $10^{-3}$.
More importantly, the normalized mean bias is most of the time smaller than the mean absolute relative error by at least an order of magnitude.
We can conclude that there is no significant bias in the predictions of the network.
We find similar results for the other combinations of statistics considered (see \ref{app:performance_test_set}).

In \cref{fig:r2score_cnn_vs_ps_test_set}, we present the $R^2$ score for each individual parameter of interest and for each model considered.
For each model considered, the best measured parameters are $\Omega_{\rm m}$, $\sigma_{8}$, and their combination $S_8$.
The CNN only model performs slightly worse than the baseline CA+PS model for the parameters $\Omega_{\rm m}$ and $\sigma_8$, and slightly better for the parameters $h$, $n_{\rm s}$, and $\Omega_{\rm b}$.
However, it performs very poorly in measuring the $M-L_{\rm X}$ scaling relation parameters.
When the cluster abundance information is included, all the CNN-based models perform better than the CA+PS model, except for the $b_{\rm X}$ and $c_{\rm X}$ parameters.
Overall, including the X-ray luminosity information in the input of the CNN improves the regression, with the weighted CA+$\mathrm{CNN}_{w_{\rm X}}$ model giving the best results.
Finally, when the power spectrum is explicitly combined with the field-level analysis, that is, the CA+CNN+PS model, the performance is similar to the CA+CNN case.

\begin{figure}
    \centering
    \includegraphics[width=8cm]{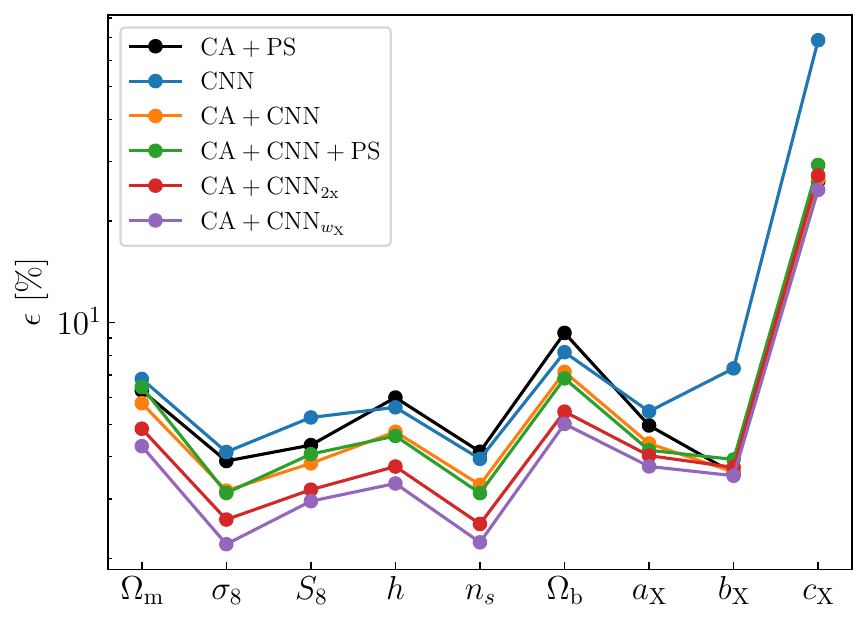}
    \caption{
        Mean absolute relative error evaluated on the test set for each target parameter.
        We have also included the derived parameter $S_8$.
        Each colour represents a different combination of statistics.
        We exclude $\sigma_{\rm tot}$, since it is unconstrained in all the cases considered.
    }
    \label{fig:mean_abs_err_cnn_vs_ps_test_set}
\end{figure}

In order to compare the performance of the different networks in extracting cosmological parameters, we now focus on the mean absolute relative error as defined in \cref{eq:mean_abs_rel_err}.
In \cref{fig:mean_abs_err_cnn_vs_ps_test_set}, we present such metric for each individual parameter and for the different networks considered.
Once again, we include the derived parameter $S_8$ and exclude $\sigma_{\rm tot}$ from the comparison.
For the cosmological parameters, all the CNN-based models joined with the cluster abundance perform better than the standard CA+PS combination.
The best performing scenario is the CNN trained on the weighted overdensity field.
The CNN with two luminosity bins as input channels gives an intermediate improvement with respect to the base CNN case, but performs worse than the weighted case.
In general, for the $M-L_{\rm X}$ scaling relation parameters, we see no significant improvement in the CNN-based scenarios with respect to the baseline CA+PS.
There is only an improvement on the parameter $a_{\rm X}$, while the parameters $b_{\rm X}$ and $c_{\rm X}$ are in some cases slightly less well constrained.
These results are in agreement with the $R^2$ score measurements presented in \cref{fig:r2score_cnn_vs_ps_test_set}.

\begin{figure}
    \centering
    \includegraphics[width=8cm]{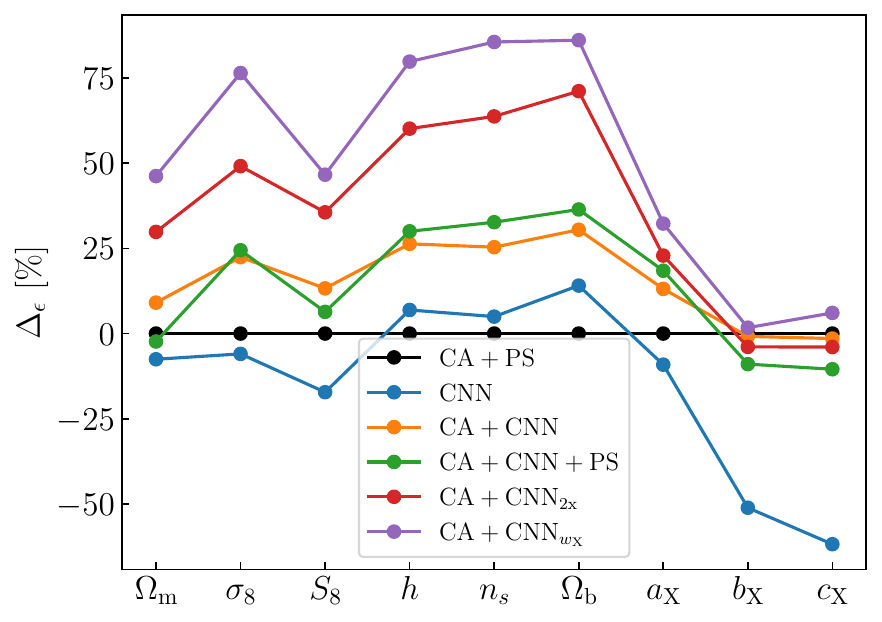}
    \caption{
        Improvement on the mean absolute relative error with respect to the reference model CA+PS, evaluated on the test set and for each target parameter.
        We have also included the derived parameter $S_8$.
        Each colour represents a different combination of statistics.
        We exclude $\sigma_{\rm tot}$, since it is unconstrained in all the cases considered.
    }
    \label{fig:frac_improv_mean_abs_err_cnn_vs_ps_test_set}
\end{figure}

In \cref{fig:frac_improv_mean_abs_err_cnn_vs_ps_test_set}, we present the same results under a different perspective, by showing the improvement on the mean absolute relative error $\epsilon$ with respect to the reference case of the CA+PS model.
We define such quantity as
\begin{equation}
    \label{eq:frac_improv_err}
    \Delta_{\epsilon} = \frac{\epsilon_{\rm ref}}{\epsilon} -1,
\end{equation}
where $\epsilon_{\rm ref}$ is the mean absolute relative error of the reference model.
The improvement with respect to the standard CA+PS case for the cosmological parameters, ranges from $\sim10\%$ to $\sim30\%$ for the CA+CNN model.
For the other CNN-based models with some X-ray luminosity information in the input overdensity field, that is, CA+$\mathrm{CNN}_{\rm 2x}$ and CA+$\mathrm{CNN}_{w_{\rm X}}$, the improvement ranges from $\sim 30\%$ to $\sim 85\%$.

\section{Discussion and conclusions}
\label{sec:conclusion}

We presented the first results of a project aimed at improving our ability to extract cosmological information from X-ray cluster survey using a machine learning field-level approach. 
Specifically, we studied the performance of convolutional neural networks when combining cluster abundance with their spatial clustering.
The main objective has been to compare the results yielded by standard summary statistics as the power spectrum, with the field-level analysis provided by CNN.

In this work we use a 3LPT code, \texttt{Pinocchio}, for the fast production of large numbers of dark matter halo catalogues.
More specifically, we have produced $4096$ mocks exploring the five-dimensional parameter space of the $\Lambda$CDM model ($\Omega_{\rm m}$, $\sigma_8$, $h$, $n_{\rm s}$, $\Omega_{\rm b}$), each one with a different set of cosmological parameters and a different seed for the initial conditions.
The volume and resolution of these simulations are designed to properly simulate cluster-sized massive haloes with masses $\geq 10^{14}\,h^{-1}\,M_{\odot}$.
X-ray luminosities were assigned to each individual halo using the empirical $M-L_{\rm X}$ scaling relation from \citet{Balaguera_2012}, producing synthetic catalogues of X-ray selected galaxy clusters.
For each cosmological model, $8$ different variants for each of the four parameters describing the $M-L_{\rm X}$ relation ($a_{\rm X}$, $b_{\rm X}$, $c_{\rm X}$, and $\sigma_{\rm tot}$) were considered, yielding a total of $32\,768$ mock cluster catalogues sampling the $(5+4)$-dimensional parameter space.

For each cosmology, we focused here on a single snapshot at $z=0.1$, considering clusters with X-ray luminosity $L_{\rm X} \geq 3 \times 10^{43}\,h^{-2}\,\mathrm{erg}\,\mathrm{s}^{-1}$.
This is consistent with the sensitivity of the REFLEX sample at $z=0.1$ \citep{boehringer04}, which is close to the median redshift of the survey.  
For the fiducial cosmology and $M-L_{\rm X}$ parameters, this yields around $14\,500$ clusters in a comoving volume of $\left(1500\,h^{-1}\,\mathrm{Mpc}\right)^3$.
In all cases, the analyses are limited within the range of scales $4.2\times10^{-3}\,h\,\mathrm{Mpc}^{-1} \lesssim k \lesssim 1.3\times10^{-1}\,h\,\mathrm{Mpc}^{-1}$.

Our results show that a CNN is capable of extracting information more efficiently than the binned power spectrum.
For the $M-L_{\rm X}$ scaling relation parameters, there is no significant improvement in using a CNN with respect to the power spectrum.
However, in the case of the cosmological parameters, the mean absolute relative error with which the neural network extracts information sees an improvement ranging from $\sim10\%$ to $\sim 30\%$, depending on the specific parameter considered.

We also found that the performance of the CNN is improved if information about the X-ray luminosity of the clusters is passed to the CNN.
We explored two options for this.
First, we computed a weighted overdensity field, built by assigning to each individual cluster a weight proportional to its luminosity.
Second, we compute two unweighted overdensity fields using two different cuts in luminosity and then combine them as input channels to the CNN.
The gain in precision on the derived cosmological parameters ranges from $\sim 30\%$ to $\sim 85\%$, with the luminosity-weighted field providing the best improvement.

An important hyperparameter for the CNN-based models, is the number of channels used in each convolution layer, which in our setup is controlled through the $N_{\rm chs\_first}$ hyperparameter (see \cref{subsec:architecture}).
For all the models considered, the hyperparameter tuning procedure finds an optimal value of $10$ that corresponds to the upper bound allowed by the imposed search range.
Allowing for larger values of $N_{\rm chs\_first}$ improves the performance of all CNN-based networks considered in this work, while also increasing the number of free parameters in the network.
As expected, this results in a computationally more expensive network, both in terms of training time and memory usage.
Additionally, there is the concern of overfitting the approximate \texttt{Pinocchio} mocks.

Tracing the exact source of the information learned by a neural network remains a recurring challenge in computer science and mathematics, as no trivial solution is available.
For clustering studies, the usual assumption is that CNNs have access to non-Gaussian information at small scales not captured by the power spectrum.
As explained in \cref{subsec:cosmo_probes}, the choice of the minimum scale we consider, defined by the cell size of the grid used both for the power spectrum computation and the CNN analysis, is guided by the impact of the shot noise on the power spectrum.
Indeed, since we are considering a rather sparse sample for smaller scales the power spectrum is dominated by shot noise.
At such quasi-linear scales, that is, $k\leq 1.3\times 10^{-1}\,h\,\mathrm{Mpc}^{-1}$, it is unclear how much higher-order information there is for the CNN to extract.
This suggests that the CNN’s improved performance may be partly due to its ability to compress two-point information more accurately than the binned power spectrum.
We have tried to limit this effect by choosing an optimum binning for the power spectrum, as detailed in \ref{app:ps_binning}.
Alternatively, it would be possible to test the CNN on smaller scales using a denser cluster sample.
This can be achieved by considering a lower X-ray luminosity limit, which in this work is set to that of the REFLEX survey.
One could also apply the methodology presented in this paper to a different class of galaxy clusters, such as optically selected ones.
We leave the exploration of such possibilities to future work.

Additional work is needed to obtain posterior distributions for the cosmological parameters based on the CNNs presented in this paper.
In fact, the metric we use to compare the performance of CNN and power spectrum is very simplistic.
A more detailed comparison, as well as a real application of the CNN to observational data, will require a full Bayesian analysis.
The usual method to obtain posterior distributions of the cosmological parameters from CNN-based compressed statistics relies on the so-called likelihood-free inference approach \citep[see e.g.][]{Marin_2012,Cranmer_2020,Jeffrey_2020,Lemos_2024,Ho_2024,Jeffrey_2025}.

Even more important, while the tests presented in this paper show that the cosmology of the \texttt{Pinocchio} mocks is recovered self-consistently by the trained CNN, there is no guarantee that its application to real data yields unbiased results.
In fact, this is what we seem to see when applying the trained CNN to independent high-resolution $N$-body simulations from the \textsc{AbacusSummit} simulation suite \citep{Maksimova_2021}.
The goal of this ongoing, final part of the project, is to develop a robust multi-fidelity framework that complements the analysis presented in this work by incorporating information from a set of high-fidelity simulations, which more accurately capture the complexity of real data and the physics behind them.
In this broader picture, the low-fidelity (\texttt{Pinocchio}) simulations are used to identify an efficient compression of the information encoded in the field distribution. 
The high-fidelity simulations, in turn, are then employed to train the extraction of cosmological parameters from this compressed statistic, thus reducing the number of high-fidelity simulations required for a fully reliable cosmological inference (Sáez-Casares et al., in preparation).

Additionally, for a more realistic application to real data, the \texttt{Pinocchio} training samples will need to include more accurate observational features, as the survey geometry, volume and selection function.
We have neglected these effects in order to work in a more controlled environment.
In this simpler, yet sufficiently realistic, setup, we have demonstrated the ability of CNN to extract additional information, compared to the traditional approach.
At the time of writing this paper, we are already experimenting the application of the same scheme to the well-known REFLEX survey data \citep{boehringer04, guzzo09}, which in terms of cluster luminosity and redshift range covered is close to the characteristics of the cubic \texttt{Pinocchio} mocks used so far.
Furthermore, with its limited size and volume, it provides a first, computationally easy benchmark for our inference pipeline, while being a robust reference in terms of classic estimates of cosmological parameters from galaxy clusters \citep{schuecker02, schuecker03a}.
The following step will be, then, to tailor the algorithm to cluster catalogues from the most up-to-date X-ray survey, i.e., that from the eROSITA satellite \citep{erosita_2024}.

\section*{Acknowledgements}
This work is supported by Italian Research Center on High Performance Computing Big Data and Quantum Computing (ICSC), project funded by European Union - NextGenerationEU - and National Recovery and Resilience Plan (NRRP) - Mission 4 Component 2 within the activities of Spoke 3 (Astrophysics and Cosmos Observations).
This project has received funding from the European Union NextGeneration EU program – NRP Mission 4 Component 2 Investment 1.1 – MUR PRIN 2022 – Code 2022SKKYJN.
Computational resources provided by INDACO Platform, which is a project of High Performance Computing at the University of MILAN (\url{http://www.unimi.it}).
The authors acknowledge the computational resources provided by CINECA through the ISCRA initiative, using the Leonardo supercomputer. 
MC and JM are partially supported by the 2024/25 ``Research and Education'' grant from Fondazione CRT. The OAVdA is managed by the Fondazione Cl\'ement Fillietroz-ONLUS, which is supported by the Regional Government of the Aosta Valley, the Town Municipality of Nus and the ``Unit\'e des Communes vald\^otaines Mont-\'Emilius''.
We are grateful to Ben Granett for spearheading the use of {\tt Pinocchio} to generate training samples for CNN analyses of galaxy survey data.
We thank Pierre Zhang for useful discussions.
This work has made used of the following numerical libraries on top of the ones already cited: \texttt{SciPy}~\citep{scipy}, \texttt{NumPy}~\citep{numpy}, \texttt{scikit-learn}~\citep{scikit-learn}, and \texttt{Matplotlib}~\citep{matplotlib}.

\appendix

\section{Impact of the power spectrum binning}
\label{app:ps_binning}

\begin{figure}
    \centering
    \includegraphics[width=\linewidth]{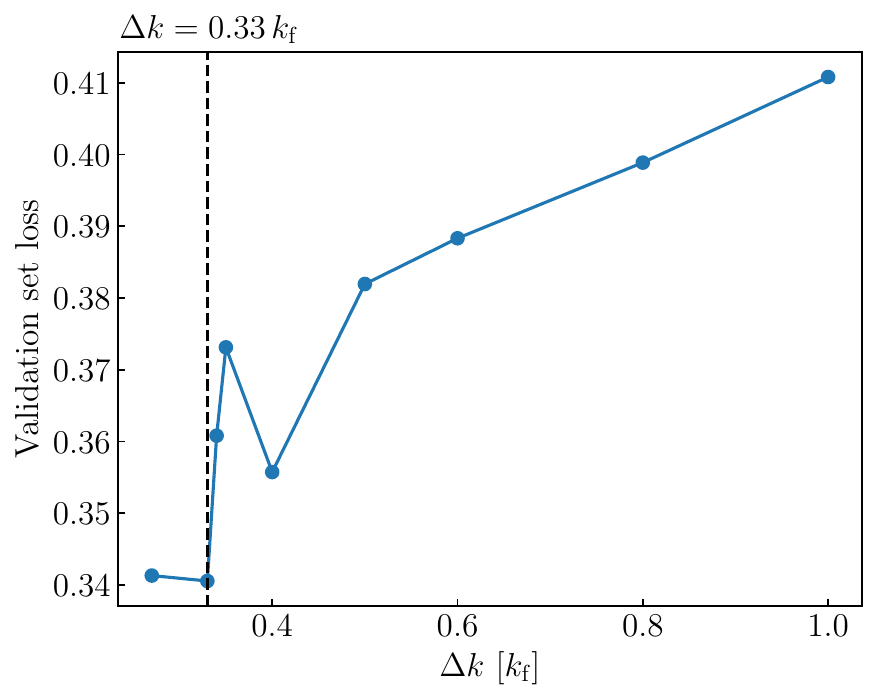}
    \caption{
        Validation set loss of the CA+PS model as a function of the bin width used for the input power spectrum in units of the fundamental wavenumber of the simulation box $k_{\rm f}$.
        The dashed line marks the optimum bin width of $\Delta k=0.33\,k_{\rm f}$.
    }
    \label{fig:ps_binning}
\end{figure}

The capacity of a neural network to extract cosmological information from a binned power spectrum is sensitive to the wavenumber binning used.
Indeed, if the power spectrum is computed with large wavenumber bins, some information of the three-dimensional power spectrum is lost.
Ideally, we would provide as input to the neural network a lossless compression of the three-dimensional power spectrum.
This would require a non-trivial binning scheme that goes beyond the scope of this paper.

We settle for a simpler strategy that still gives us meaningful results.
We bin the power spectrum using constant linearly spaced bins of width $\Delta k$.
We want to find the value of $\Delta k$ that allows the neural network to obtain the best values for the cosmological parameters.
In some sense, the parameter $\Delta k$ could be seen as an hyperparameter of the network and optimised with \texttt{optuna} at the same time as the others.
However, since training the CA+PS model is fast, we follow a simpler procedure.
We consider different value of $\Delta k$ and each time repeat the full hyperparameter optimization as described in \cref{subsec:hp_optim}.
We then consider the best loss obtained on the validation set as a metric to compare different bin widths.

In \cref{fig:ps_binning} we show the best validation loss as a function of the bin width $\Delta k$ in units of the fundamental mode of the simulation box $k_{\rm f}$.
We can see that the commonly used value of the power spectrum bin width equal to $k_{\rm f}$ does not produce the best results.
The neural network is able to perform a better regression of the cosmological parameters with $\Delta k=0.33\,k_{\rm f}$.
We note that when the binning becomes small with respect to $k_{\rm f}$ the training of the neural network becomes unstable.

\section{Test set performance}
\label{app:performance_test_set}

\begin{figure*}
    \centering
    \includegraphics[width=\linewidth]{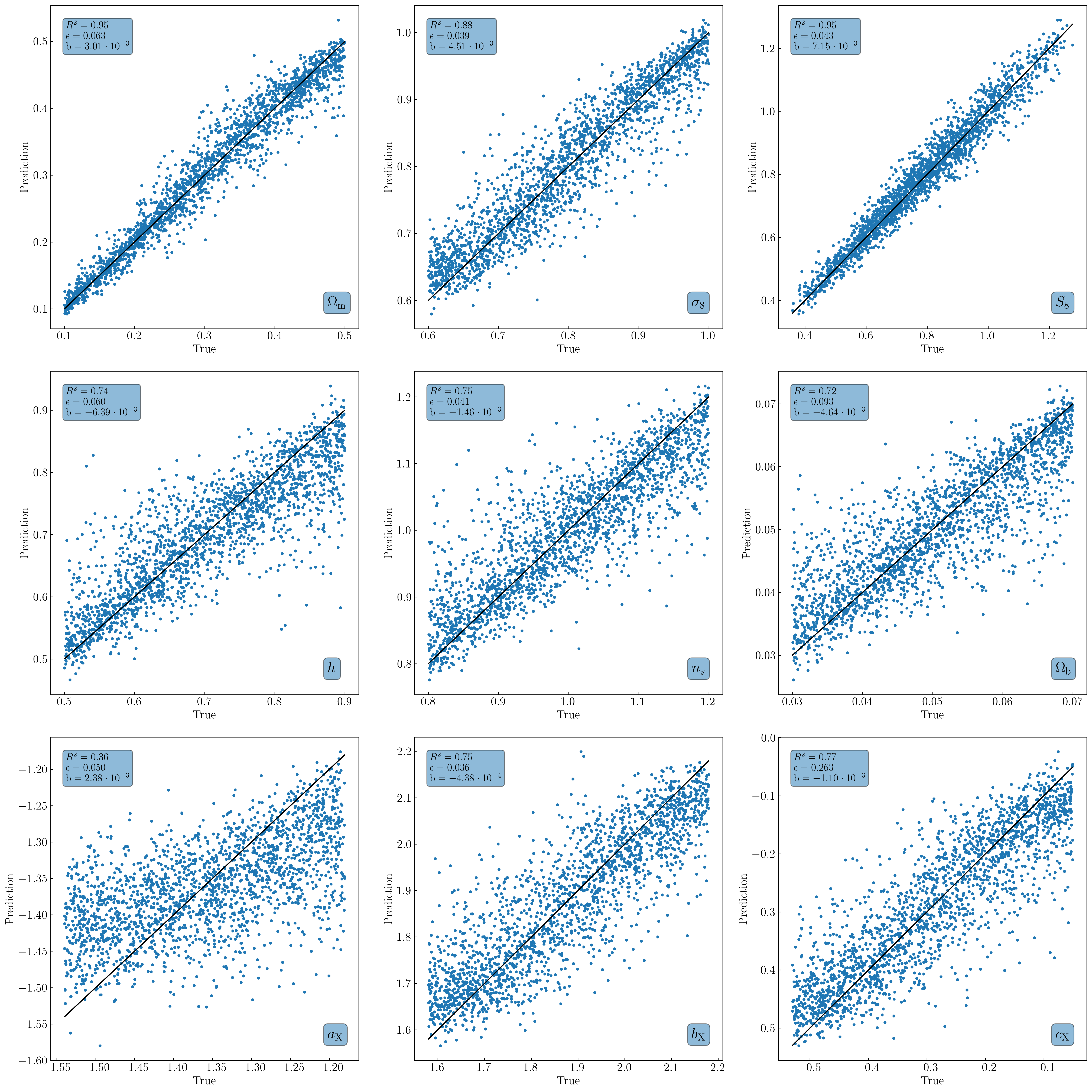}
    \caption{
        Same as \cref{fig:test_set_eval_mosaic_cnn} for the CA+PS model.
    }
    \label{fig:test_set_eval_mosaic_ps}
\end{figure*}

\begin{figure*}
    \centering
    \includegraphics[width=\linewidth]{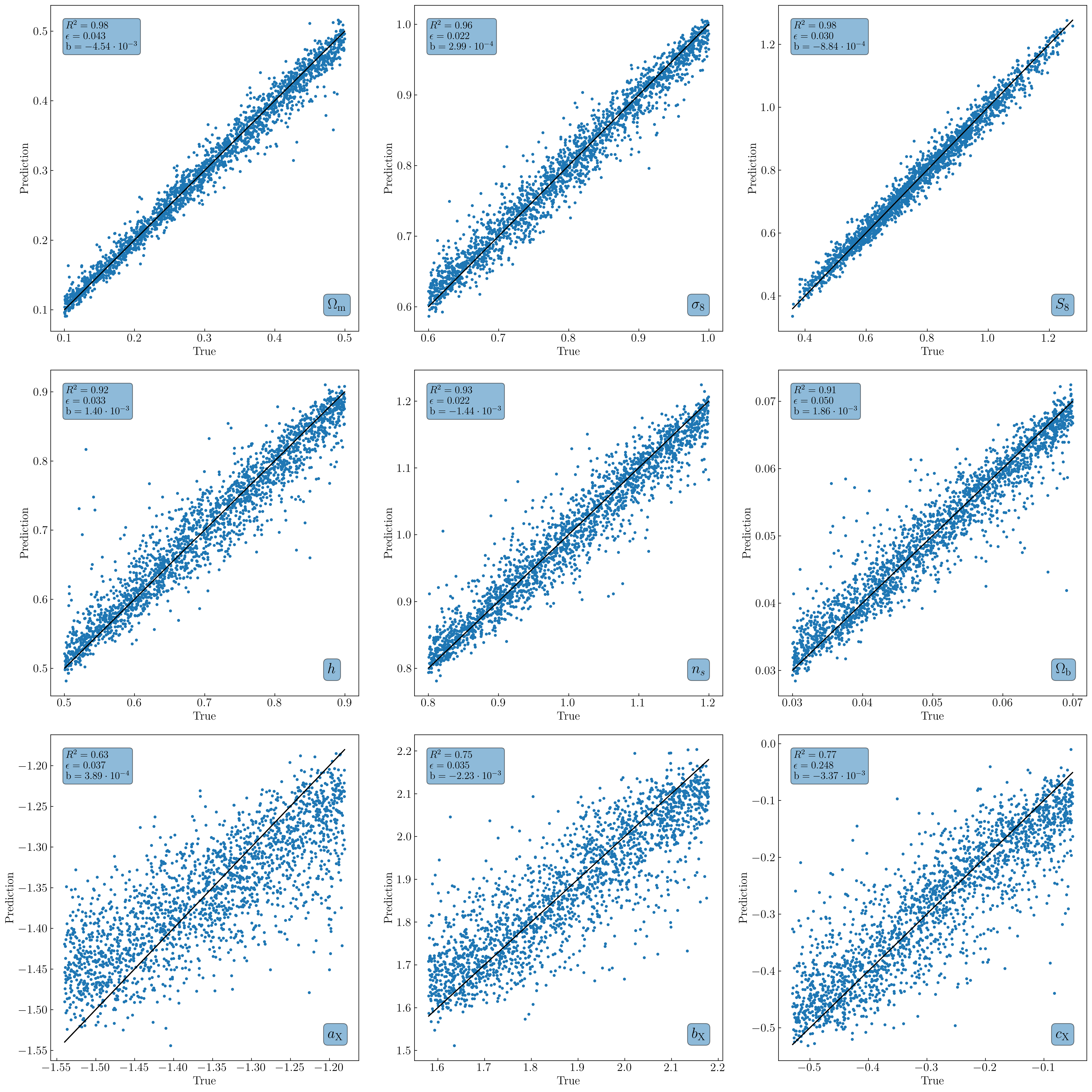}
    \caption{
        Same as \cref{fig:test_set_eval_mosaic_cnn} for the CA+$\mathrm{CNN}_{w_{\rm X}}$ model.
    }
    \label{fig:test_set_eval_mosaic_cnn_m}
\end{figure*}

\begin{figure*}
    \centering
    \includegraphics[width=\linewidth]{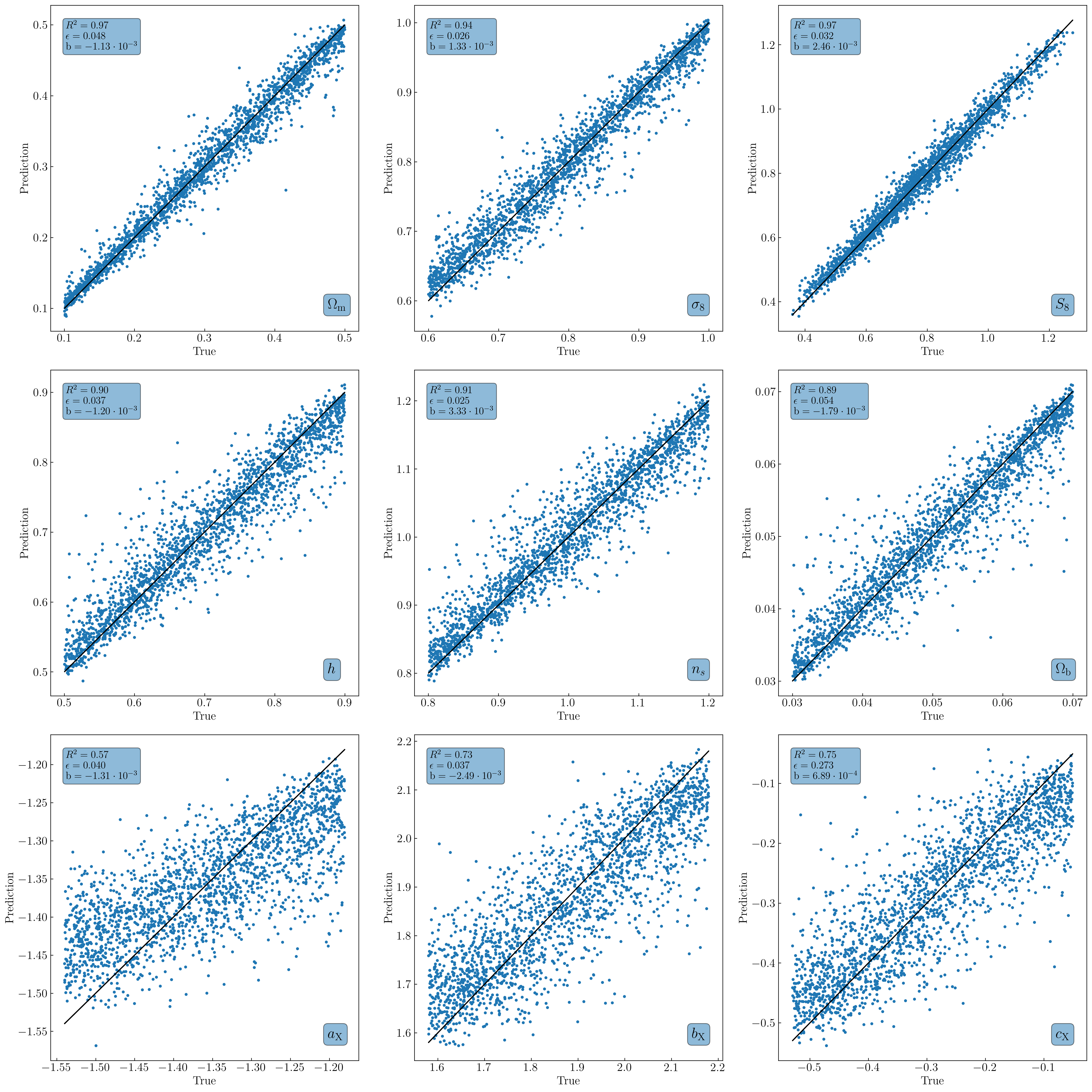}
    \caption{
        Same as \cref{fig:test_set_eval_mosaic_cnn} for the CA+$\mathrm{CNN}_{\rm 2x}$ model.
    }
    \label{fig:test_set_eval_mosaic_cnn_2x}
\end{figure*}

\begin{figure*}
    \centering
    \includegraphics[width=\linewidth]{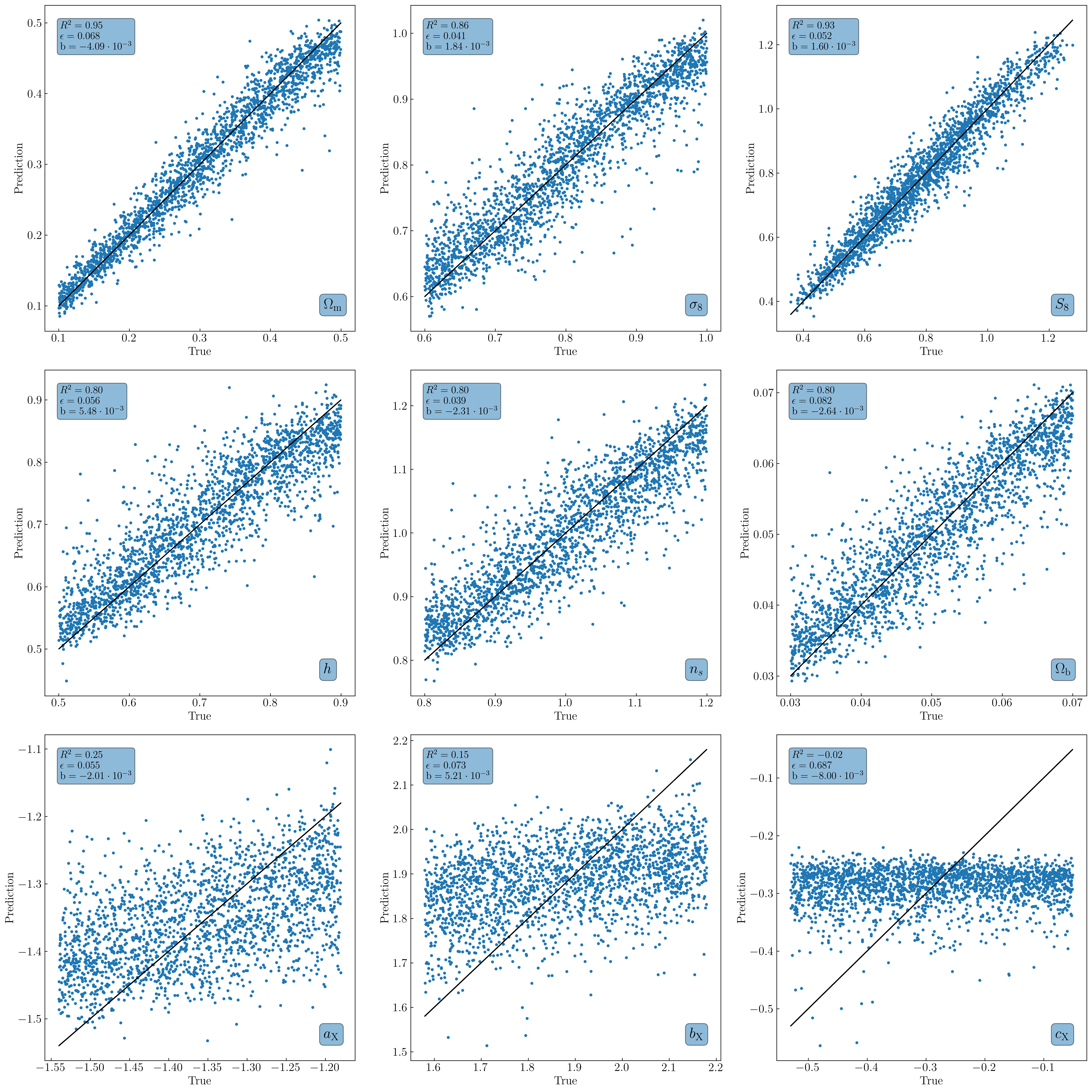}
    \caption{
        Same as \cref{fig:test_set_eval_mosaic_cnn} for the CNN model.
    }
    \label{fig:test_set_eval_mosaic_cnn_only}
\end{figure*}

\begin{figure*}
    \centering
    \includegraphics[width=\linewidth]{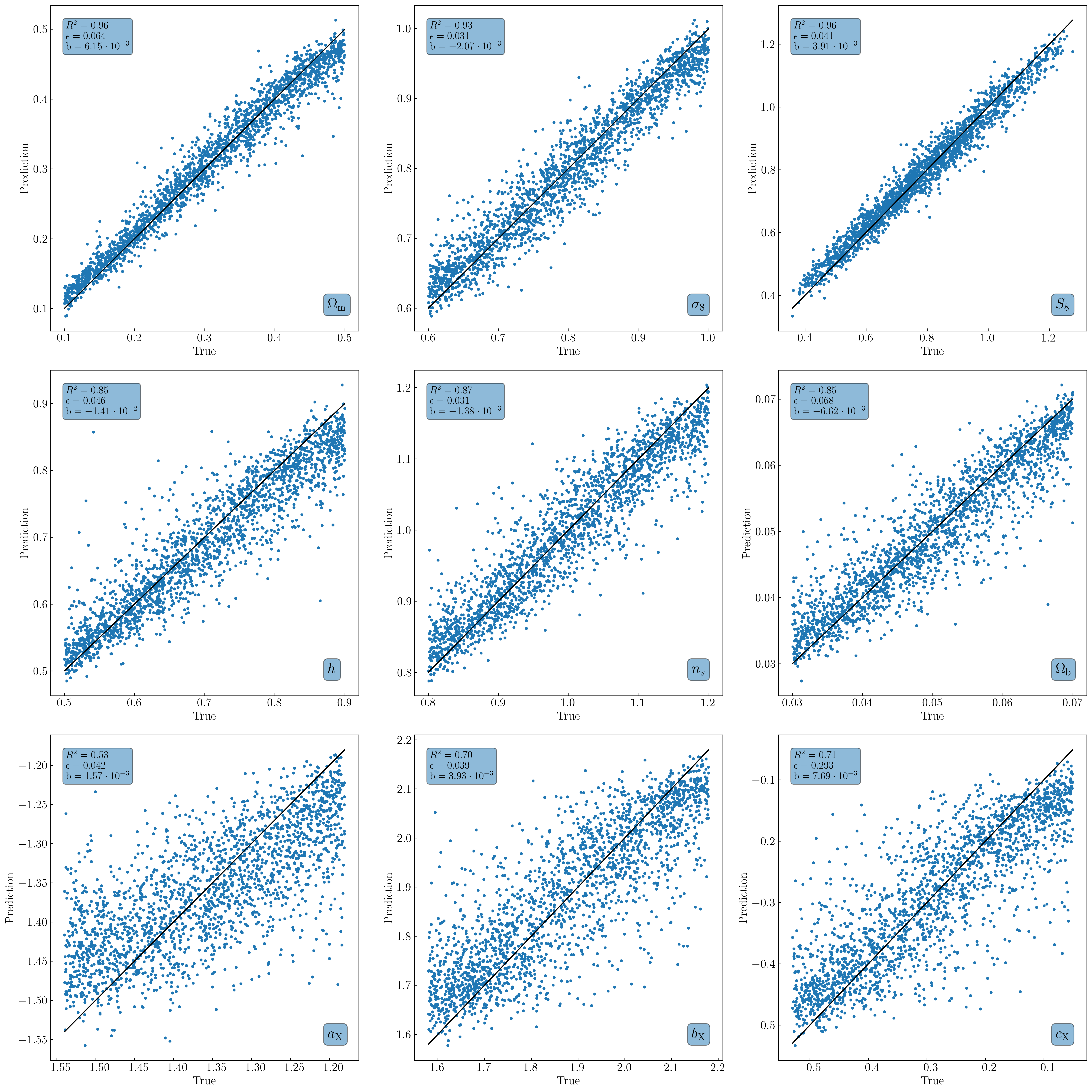}
    \caption{
        Same as \cref{fig:test_set_eval_mosaic_cnn} for the CA+CNN+PS model.
    }
    \label{fig:test_set_eval_mosaic_cnn_ps_nc}
\end{figure*}

In this appendix, we show the performance plots for the other models considered in this work not shown in \cref{sec:results} (see \cref{fig:test_set_eval_mosaic_ps,fig:test_set_eval_mosaic_cnn_m,fig:test_set_eval_mosaic_cnn_2x,fig:test_set_eval_mosaic_cnn_only,fig:test_set_eval_mosaic_cnn_ps_nc}).

\bibliographystyle{elsarticle-harv-mod} 
\bibliography{bibliography}

@article{PlanckXXVII_SZ_2025,
	adsurl = {https://ui.adsabs.harvard.edu/abs/2016A&A...594A..27P},
	archiveprefix = {arXiv},
	author = {{Planck Collaboration} and {Ade}, P.~A.~R. and {Aghanim}, N. and {Arnaud}, M. and {Ashdown}, M. and {Aumont}, J. and {Baccigalupi}, C. and {Banday}, A.~J. and {Barreiro}, R.~B. and {Barrena}, R. and {Bartlett}, J.~G. and {Bartolo}, N. and {Battaner}, E. and {Battye}, R. and {Benabed}, K. and {Beno{\^\i}t}, A. and {Benoit-L{\'e}vy}, A. and {Bernard}, J.-P. and {Bersanelli}, M. and {Bielewicz}, P. and {Bikmaev}, I. and {B{\"o}hringer}, H. and {Bonaldi}, A. and {Bonavera}, L. and {Bond}, J.~R. and {Borrill}, J. and {Bouchet}, F.~R. and {Bucher}, M. and {Burenin}, R. and {Burigana}, C. and {Butler}, R.~C. and {Calabrese}, E. and {Cardoso}, J.-F. and {Carvalho}, P. and {Catalano}, A. and {Challinor}, A. and {Chamballu}, A. and {Chary}, R.-R. and {Chiang}, H.~C. and {Chon}, G. and {Christensen}, P.~R. and {Clements}, D.~L. and {Colombi}, S. and {Colombo}, L.~P.~L. and {Combet}, C. and {Comis}, B. and {Couchot}, F. and {Coulais}, A. and {Crill}, B.~P. and {Curto}, A. and {Cuttaia}, F. and {Dahle}, H. and {Danese}, L. and {Davies}, R.~D. and {Davis}, R.~J. and {de Bernardis}, P. and {de Rosa}, A. and {de Zotti}, G. and {Delabrouille}, J. and {D{\'e}sert}, F.-X. and {Dickinson}, C. and {Diego}, J.~M. and {Dolag}, K. and {Dole}, H. and {Donzelli}, S. and {Dor{\'e}}, O. and {Douspis}, M. and {Ducout}, A. and {Dupac}, X. and {Efstathiou}, G. and {Eisenhardt}, P.~R.~M. and {Elsner}, F. and {En{\ss}lin}, T.~A. and {Eriksen}, H.~K. and {Falgarone}, E. and {Fergusson}, J. and {Feroz}, F. and {Ferragamo}, A. and {Finelli}, F. and {Forni}, O. and {Frailis}, M. and {Fraisse}, A.~A. and {Franceschi}, E. and {Frejsel}, A. and {Galeotta}, S. and {Galli}, S. and {Ganga}, K. and {G{\'e}nova-Santos}, R.~T. and {Giard}, M. and {Giraud-H{\'e}raud}, Y. and {Gjerl{\o}w}, E. and {Gonz{\'a}lez-Nuevo}, J. and {G{\'o}rski}, K.~M. and {Grainge}, K.~J.~B. and {Gratton}, S. and {Gregorio}, A. and {Gruppuso}, A. and {Gudmundsson}, J.~E. and {Hansen}, F.~K. and {Hanson}, D. and {Harrison}, D.~L. and {Hempel}, A. and {Henrot-Versill{\'e}}, S. and {Hern{\'a}ndez-Monteagudo}, C. and {Herranz}, D. and {Hildebrandt}, S.~R. and {Hivon}, E. and {Hobson}, M. and {Holmes}, W.~A. and {Hornstrup}, A. and {Hovest}, W. and {Huffenberger}, K.~M. and {Hurier}, G. and {Jaffe}, A.~H. and {Jaffe}, T.~R. and {Jin}, T. and {Jones}, W.~C. and {Juvela}, M. and {Keih{\"a}nen}, E. and {Keskitalo}, R. and {Khamitov}, I. and {Kisner}, T.~S. and {Kneissl}, R. and {Knoche}, J. and {Kunz}, M. and {Kurki-Suonio}, H. and {Lagache}, G. and {Lamarre}, J.-M. and {Lasenby}, A. and {Lattanzi}, M. and {Lawrence}, C.~R. and {Leonardi}, R. and {Lesgourgues}, J. and {Levrier}, F. and {Liguori}, M. and {Lilje}, P.~B. and {Linden-V{\o}rnle}, M. and {L{\'o}pez-Caniego}, M. and {Lubin}, P.~M. and {Mac{\'\i}as-P{\'e}rez}, J.~F. and {Maggio}, G. and {Maino}, D. and {Mak}, D.~S.~Y. and {Mandolesi}, N. and {Mangilli}, A. and {Martin}, P.~G. and {Mart{\'\i}nez-Gonz{\'a}lez}, E. and {Masi}, S. and {Matarrese}, S. and {Mazzotta}, P. and {McGehee}, P. and {Mei}, S. and {Melchiorri}, A. and {Melin}, J.-B. and {Mendes}, L. and {Mennella}, A. and {Migliaccio}, M. and {Mitra}, S. and {Miville-Desch{\^e}nes}, M.-A. and {Moneti}, A. and {Montier}, L. and {Morgante}, G. and {Mortlock}, D. and {Moss}, A. and {Munshi}, D. and {Murphy}, J.~A. and {Naselsky}, P. and {Nastasi}, A. and {Nati}, F. and {Natoli}, P. and {Netterfield}, C.~B. and {N{\o}rgaard-Nielsen}, H.~U. and {Noviello}, F. and {Novikov}, D. and {Novikov}, I. and {Olamaie}, M. and {Oxborrow}, C.~A. and {Paci}, F. and {Pagano}, L. and {Pajot}, F. and {Paoletti}, D. and {Pasian}, F. and {Patanchon}, G. and {Pearson}, T.~J. and {Perdereau}, O. and {Perotto}, L. and {Perrott}, Y.~C. and {Perrotta}, F. and {Pettorino}, V. and {Piacentini}, F. and {Piat}, M. and {Pierpaoli}, E. and {Pietrobon}, D. and {Plaszczynski}, S. and {Pointecouteau}, E. and {Polenta}, G. and {Pratt}, G.~W. and {Pr{\'e}zeau}, G. and {Prunet}, S. and {Puget}, J.-L.},
	doi = {10.1051/0004-6361/201525823},
	eid = {A27},
	eprint = {1502.01598},
	journal = {A\&A},
	month = sep,
	pages = {A27},
	primaryclass = {astro-ph.CO},
	title = {{Planck 2015 results. XXVII. The second Planck catalogue of Sunyaev-Zeldovich sources}},
	volume = {594},
	year = 2016}

@article{SPT_SZ_Clusters_2015,
	adsurl = {https://ui.adsabs.harvard.edu/abs/2015ApJS..216...27B},
	archiveprefix = {arXiv},
	author = {{Bleem}, L.~E. and {Stalder}, B. and {de Haan}, T. and {Aird}, K.~A. and {Allen}, S.~W. and {Applegate}, D.~E. and {Ashby}, M.~L.~N. and {Bautz}, M. and {Bayliss}, M. and {Benson}, B.~A. and {Bocquet}, S. and {Brodwin}, M. and {Carlstrom}, J.~E. and {Chang}, C.~L. and {Chiu}, I. and {Cho}, H.~M. and {Clocchiatti}, A. and {Crawford}, T.~M. and {Crites}, A.~T. and {Desai}, S. and {Dietrich}, J.~P. and {Dobbs}, M.~A. and {Foley}, R.~J. and {Forman}, W.~R. and {George}, E.~M. and {Gladders}, M.~D. and {Gonzalez}, A.~H. and {Halverson}, N.~W. and {Hennig}, C. and {Hoekstra}, H. and {Holder}, G.~P. and {Holzapfel}, W.~L. and {Hrubes}, J.~D. and {Jones}, C. and {Keisler}, R. and {Knox}, L. and {Lee}, A.~T. and {Leitch}, E.~M. and {Liu}, J. and {Lueker}, M. and {Luong-Van}, D. and {Mantz}, A. and {Marrone}, D.~P. and {McDonald}, M. and {McMahon}, J.~J. and {Meyer}, S.~S. and {Mocanu}, L. and {Mohr}, J.~J. and {Murray}, S.~S. and {Padin}, S. and {Pryke}, C. and {Reichardt}, C.~L. and {Rest}, A. and {Ruel}, J. and {Ruhl}, J.~E. and {Saliwanchik}, B.~R. and {Saro}, A. and {Sayre}, J.~T. and {Schaffer}, K.~K. and {Schrabback}, T. and {Shirokoff}, E. and {Song}, J. and {Spieler}, H.~G. and {Stanford}, S.~A. and {Staniszewski}, Z. and {Stark}, A.~A. and {Story}, K.~T. and {Stubbs}, C.~W. and {Vanderlinde}, K. and {Vieira}, J.~D. and {Vikhlinin}, A. and {Williamson}, R. and {Zahn}, O. and {Zenteno}, A.},
	doi = {10.1088/0067-0049/216/2/27},
	eid = {27},
	eprint = {1409.0850},
	journal = {ApJS},
	month = feb,
	number = {2},
	pages = {27},
	primaryclass = {astro-ph.CO},
	title = {{Galaxy Clusters Discovered via the Sunyaev-Zel'dovich Effect in the 2500-Square-Degree SPT-SZ Survey}},
	volume = {216},
	year = 2015}

@article{ACT_SZ_Clusters_2025,
	adsurl = {https://ui.adsabs.harvard.edu/abs/2025arXiv250721459A},
	archiveprefix = {arXiv},
	author = {{ACTDESHSC Collaboration} and {Aguena}, M. and {Aiola}, S. and {Allam}, S. and {Andrade-Oliveira}, F. and {Bacon}, D. and {Bahcall}, N. and {Battaglia}, N. and {Battistelli}, E.~S. and {Bocquet}, S. and {Bolliet}, B. and {Bond}, J.~R. and {Brooks}, D. and {Calabrese}, E. and {Carretero}, J. and {Choi}, S.~K. and {da Costa}, L.~N. and {Costanzi}, M. and {Coulton}, W. and {Davis}, T.~M. and {Desai}, S. and {Devlin}, M.~J. and {Dicker}, S. and {Doel}, P. and {Duivenvoorden}, A.~J. and {Dunkley}, J. and {Ferraro}, S. and {Flaugher}, B. and {Frieman}, J. and {Gallardo}, P.~A. and {Gatti}, M. and {Gaztanaga}, E. and {Gill}, A.~S. and {Golec}, J.~E. and {Gruen}, D. and {Gruendl}, R.~A. and {Halpern}, M. and {Hasselfield}, M. and {Hill}, J.~C. and {Hilton}, M. and {Hincks}, A.~D. and {Hinton}, S.~R. and {Hollowood}, D.~L. and {Honscheid}, K. and {Hubmayr}, J. and {Huffenberger}, K.~M. and {Hughes}, J.~P. and {James}, D.~J. and {Klein}, M. and {Knowles}, K. and {Koopman}, B.~J. and {Kosowsky}, A. and {Lahav}, O. and {Lee}, E. and {Lin}, Y. and {Lokken}, M. and {Madhavacheril}, M.~S. and {Plazas Malag{\'o}n}, A.~A. and {Marrewijk}, J. v. and {Marshall}, J.~L. and {McMahon}, J. and {Mena-Fern{\'a}ndez}, J. and {Miquel}, R. and {Miyatake}, H. and {Mohr}, J.~J. and {Moodley}, K. and {Mroczkowski}, T. and {Naess}, S. and {Nati}, F. and {Nicola}, A. and {Niemack}, M.~D. and {Ogando}, R.~L.~C. and {Oguri}, M. and {Orlowski-Scherer}, J. and {Page}, L.~A. and {Partridge}, B. and {da Silva Pereira}, M.~E. and {Porredon}, A. and {Qu}, F.~J. and {Ragavan}, D.~C. and {Ried Guachalla}, B. and {Romer}, A.~K. and {Carnero Rosell}, A. and {Rykoff}, E.~S. and {Samuroff}, S. and {Sanchez}, E. and {Sevilla-Noarbe}, I. and {Sierra}, C. and {Sif{\'o}n}, C. and {Smith}, M. and {Staggs}, S.~T. and {Suchyta}, E. and {Swanson}, M.~E.~C. and {Tucker}, D.~L. and {Vargas}, C. and {Vavagiakis}, E.~M. and {De Vicente}, J. and {Weaverdyck}, N. and {Weller}, J. and {Wollack}, E.~J. and {Zubeldia}, I.},
	doi = {10.48550/arXiv.2507.21459},
	eprint = {2507.21459},
	journal = {arXiv e-prints},
	month = jul,
	primaryclass = {astro-ph.CO},
	title = {{The Atacama Cosmology Telescope: DR6 Sunyaev-Zel'dovich Selected Galaxy Clusters Catalog}},
	year = 2025}

@article{Abell_1989,
	adsurl = {https://ui.adsabs.harvard.edu/abs/1989ApJS...70....1A},
	author = {{Abell}, George O. and {Corwin}, Jr., Harold G. and {Olowin}, Ronald P.},
	doi = {10.1086/191333},
	journal = {ApJS},
	month = may,
	pages = {1},
	title = {{A Catalog of Rich Clusters of Galaxies}},
	volume = {70},
	year = 1989}

@article{Abell_1958,
	adsurl = {https://ui.adsabs.harvard.edu/abs/1958ApJS....3..211A},
	author = {{Abell}, George O.},
	doi = {10.1086/190036},
	journal = {ApJS},
	month = may,
	pages = {211},
	title = {{The Distribution of Rich Clusters of Galaxies.}},
	volume = {3},
	year = 1958}

@article{Sunyaev_1972,
	adsurl = {https://ui.adsabs.harvard.edu/abs/1972CoASP...4..173S},
	author = {{Sunyaev}, R.~A. and {Zeldovich}, Ya. B.},
	journal = {Comments on Astrophysics and Space Physics},
	month = nov,
	pages = {173},
	title = {{The Observations of Relic Radiation as a Test of the Nature of X-Ray Radiation from the Clusters of Galaxies}},
	volume = {4},
	year = 1972}

@article{Fumagalli_2024,
	adsurl = {https://ui.adsabs.harvard.edu/abs/2024A&A...682A.148F},
	archiveprefix = {arXiv},
	author = {{Fumagalli}, A. and {Costanzi}, M. and {Saro}, A. and {Castro}, T. and {Borgani}, S.},
	doi = {10.1051/0004-6361/202348296},
	eid = {A148},
	eprint = {2310.09146},
	journal = {A\&A},
	month = feb,
	pages = {A148},
	primaryclass = {astro-ph.CO},
	title = {{Cosmological constraints from the abundance, weak lensing, and clustering of galaxy clusters: Application to the SDSS}},
	volume = {682},
	year = 2024}

@article{Kravtsov_2012,
	adsurl = {https://ui.adsabs.harvard.edu/abs/2012ARA&A..50..353K},
	archiveprefix = {arXiv},
	author = {{Kravtsov}, Andrey V. and {Borgani}, Stefano},
	doi = {10.1146/annurev-astro-081811-125502},
	eprint = {1205.5556},
	journal = {ARA\&A},
	month = sep,
	pages = {353-409},
	primaryclass = {astro-ph.CO},
	title = {{Formation of Galaxy Clusters}},
	volume = {50},
	year = 2012}

@incollection{Clerc_2023,
	adsurl = {https://ui.adsabs.harvard.edu/abs/2023hxga.book..123C},
	author = {{Clerc}, Nicolas and {Finoguenov}, Alexis},
	booktitle = {Handbook of X-ray and Gamma-ray Astrophysics},
	doi = {10.1007/978-981-16-4544-0_117-1},
	eid = {123},
	pages = {123},
	title = {{X-Ray Cluster Cosmology}},
	year = 2023}

@article{Ghirardini_2024,
	adsurl = {https://ui.adsabs.harvard.edu/abs/2024A&A...689A.298G},
	archiveprefix = {arXiv},
	author = {{Ghirardini}, V. and {Bulbul}, E. and {Artis}, E. and {Clerc}, N. and {Garrel}, C. and {Grandis}, S. and {Kluge}, M. and {Liu}, A. and {Bahar}, Y.~E. and {Balzer}, F. and {Chiu}, I. and {Comparat}, J. and {Gruen}, D. and {Kleinebreil}, F. and {Krippendorf}, S. and {Merloni}, A. and {Nandra}, K. and {Okabe}, N. and {Pacaud}, F. and {Predehl}, P. and {Ramos-Ceja}, M.~E. and {Reiprich}, T.~H. and {Sanders}, J.~S. and {Schrabback}, T. and {Seppi}, R. and {Zelmer}, S. and {Zhang}, X. and {Bornemann}, W. and {Brunner}, H. and {Burwitz}, V. and {Coutinho}, D. and {Dennerl}, K. and {Freyberg}, M. and {Friedrich}, S. and {Gaida}, R. and {Gueguen}, A. and {Haberl}, F. and {Kink}, W. and {Lamer}, G. and {Li}, X. and {Liu}, T. and {Maitra}, C. and {Meidinger}, N. and {Mueller}, S. and {Miyatake}, H. and {Miyazaki}, S. and {Robrade}, J. and {Schwope}, A. and {Stewart}, I.},
	doi = {10.1051/0004-6361/202348852},
	eid = {A298},
	eprint = {2402.08458},
	journal = {A\&A},
	month = sep,
	pages = {A298},
	primaryclass = {astro-ph.CO},
	title = {{The SRG/eROSITA all-sky survey: Cosmology constraints from cluster abundances in the western Galactic hemisphere}},
	volume = {689},
	year = 2024}

@article{Artis_2025,
	adsurl = {https://ui.adsabs.harvard.edu/abs/2025A&A...696A...5A},
	archiveprefix = {arXiv},
	author = {{Artis}, E. and {Bulbul}, E. and {Grandis}, S. and {Ghirardini}, V. and {Clerc}, N. and {Seppi}, R. and {Comparat}, J. and {Cataneo}, M. and {von der Linden}, A. and {Bahar}, Y.~E. and {Balzer}, F. and {Chiu}, I. and {Gruen}, D. and {Kleinebreil}, F. and {Kluge}, M. and {Krippendorf}, S. and {Li}, X. and {Liu}, A. and {Malavasi}, N. and {Merloni}, A. and {Miyatake}, H. and {Miyazaki}, S. and {Nandra}, K. and {Okabe}, N. and {Pacaud}, F. and {Predehl}, P. and {Ramos-Ceja}, M.~E. and {Reiprich}, T.~H. and {Sanders}, J.~S. and {Schrabback}, T. and {Zelmer}, S. and {Zhang}, X.},
	doi = {10.1051/0004-6361/202452584},
	eid = {A5},
	eprint = {2410.09499},
	journal = {A\&A},
	month = apr,
	pages = {A5},
	primaryclass = {astro-ph.CO},
	title = {{The SRG/eROSITA All-Sky Survey: Constraints on the structure growth from cluster number counts}},
	volume = {696},
	year = 2025}

@article{boehringer14,
	adsurl = {https://ui.adsabs.harvard.edu/abs/2014A&A...570A..31B},
	archiveprefix = {arXiv},
	author = {{B{\"o}hringer}, Hans and {Chon}, Gayoung and {Collins}, Chris A.},
	doi = {10.1051/0004-6361/201323155},
	eid = {A31},
	eprint = {1403.2927},
	journal = {A\&A},
	month = oct,
	pages = {A31},
	primaryclass = {astro-ph.CO},
	title = {{The extended ROSAT-ESO Flux Limited X-ray Galaxy Cluster Survey (REFLEX II). IV. X-ray luminosity function and first constraints on cosmological parameters}},
	volume = {570},
	year = 2014}

@article{boehringer02,
	adsurl = {https://ui.adsabs.harvard.edu/abs/2002ApJ...566...93B},
	author = {{B{\"o}hringer}, H. and {Collins}, C.~A. and {Guzzo}, L. and {Schuecker}, P. and {Voges}, W. and {Neumann}, D.~M. and {Schindler}, S. and {Chincarini}, G. and {De Grandi}, S. and {Cruddace}, R.~G. and {Edge}, A.~C. and {Reiprich}, T.~H. and {Shaver}, P.},
	doi = {10.1086/338072},
	journal = {ApJ},
	month = feb,
	number = {1},
	pages = {93-102},
	title = {{The ROSAT-ESO Flux-limited X-Ray (REFLEX) Galaxy Cluster Survey. IV. The X-Ray Luminosity Function}},
	volume = {566},
	year = 2002}

@article{Rosati_2002,
	adsurl = {https://ui.adsabs.harvard.edu/abs/2002ARA&A..40..539R},
	archiveprefix = {arXiv},
	author = {{Rosati}, Piero and {Borgani}, Stefano and {Norman}, Colin},
	doi = {10.1146/annurev.astro.40.120401.150547},
	eprint = {astro-ph/0209035},
	journal = {ARA\&A},
	month = jan,
	pages = {539-577},
	primaryclass = {astro-ph},
	title = {{The Evolution of X-ray Clusters of Galaxies}},
	volume = {40},
	year = 2002}

@article{guzzo09,
	adsurl = {https://ui.adsabs.harvard.edu/abs/2009A&A...499..357G},
	archiveprefix = {arXiv},
	author = {{Guzzo}, L. and {Schuecker}, P. and {B{\"o}hringer}, H. and {Collins}, C.~A. and {Ortiz-Gil}, A. and {de Grandi}, S. and {Edge}, A.~C. and {Neumann}, D.~M. and {Schindler}, S. and {Altucci}, C. and {Shaver}, P.~A.},
	doi = {10.1051/0004-6361/200810838},
	eprint = {0907.5457},
	journal = {A\&A},
	month = may,
	number = {2},
	pages = {357-369},
	primaryclass = {astro-ph.CO},
	title = {{The REFLEX galaxy cluster survey. VIII. Spectroscopic observations and optical atlas,}},
	volume = {499},
	year = 2009}

@article{schuecker02,
	adsurl = {https://ui.adsabs.harvard.edu/abs/2002MNRAS.335..807S},
	archiveprefix = {arXiv},
	author = {{Schuecker}, Peter and {Guzzo}, Luigi and {Collins}, Chris A. and {B{\"o}hringer}, Hans},
	doi = {10.1046/j.1365-8711.2002.05668.x},
	eprint = {astro-ph/0205342},
	journal = {MNRAS},
	month = sep,
	number = {3},
	pages = {807-816},
	primaryclass = {astro-ph},
	title = {{The ROSAT-ESO Flux-Limited X-ray (REFLEX) galaxy cluster survey - VI. Constraints on the cosmic matter density from the KL power spectrum}},
	volume = {335},
	year = 2002}

@article{schuecker03a,
	adsurl = {https://ui.adsabs.harvard.edu/abs/2003A&A...398..867S},
	archiveprefix = {arXiv},
	author = {{Schuecker}, P. and {B{\"o}hringer}, H. and {Collins}, C.~A. and {Guzzo}, L.},
	doi = {10.1051/0004-6361:20021715},
	eprint = {astro-ph/0208251},
	journal = {A\&A},
	month = feb,
	pages = {867-877},
	primaryclass = {astro-ph},
	title = {{The REFLEX galaxy cluster survey. VII. Omega$_{m}$ and sigma$_{8}$ from cluster abundance and large-scale clustering}},
	volume = {398},
	year = 2003}

@article{boehringer04,
	adsurl = {https://ui.adsabs.harvard.edu/abs/2004A&A...425..367B},
	archiveprefix = {arXiv},
	author = {{B{\"o}hringer}, H. and {Schuecker}, P. and {Guzzo}, L. and {Collins}, C.~A. and {Voges}, W. and {Cruddace}, R.~G. and {Ortiz-Gil}, A. and {Chincarini}, G. and {De Grandi}, S. and {Edge}, A.~C. and {MacGillivray}, H.~T. and {Neumann}, D.~M. and {Schindler}, S. and {Shaver}, P.},
	doi = {10.1051/0004-6361:20034484},
	eprint = {astro-ph/0405546},
	journal = {A\&A},
	month = oct,
	pages = {367-383},
	primaryclass = {astro-ph},
	title = {{The ROSAT-ESO Flux Limited X-ray (REFLEX) Galaxy cluster survey. V. The cluster catalogue}},
	volume = {425},
	year = 2004}

@article{collins00,
	adsurl = {https://ui.adsabs.harvard.edu/abs/2000MNRAS.319..939C},
	archiveprefix = {arXiv},
	author = {{Collins}, C.~A. and {Guzzo}, L. and {B{\"o}hringer}, H. and {Schuecker}, P. and {Chincarini}, G. and {Cruddace}, R. and {De Grandi}, S. and {MacGillivray}, H.~T. and {Neumann}, D.~M. and {Schindler}, S. and {Shaver}, P. and {Voges}, W.},
	doi = {10.1046/j.1365-8711.2000.03918.x},
	eprint = {astro-ph/0008245},
	journal = {MNRAS},
	month = dec,
	number = {3},
	pages = {939-948},
	primaryclass = {astro-ph},
	title = {{The ROSAT-ESO Flux-Limited X-ray (REFLEX) galaxy cluster survey - II. The spatial correlation function}},
	volume = {319},
	year = 2000}

@article{Lewis_1999,
	author = {Lewis, Antony and Challinor, Anthony and Lasenby, Anthony},
	doi = {10.1086/309179},
	journal = {ApJ},
	pages = {473-476},
	title = {{Efficient computation of CMB anisotropies in closed FRW models}},
	volume = {538},
	year = {2000}}

@article{Monaco_2002,
	author = {Monaco, Pierluigi and Theuns, Tom and Taffoni, Giuliano},
	doi = {10.1046/j.1365-8711.2002.05162.x},
	journal = {MNRAS},
	month = {04},
	number = {3},
	pages = {587-608},
	title = {The pinocchio algorithm: pinpointing orbit-crossing collapsed hierarchical objects in a linear density field},
	volume = {331},
	year = {2002}}

@inproceedings{optuna_2019,
	author = {Akiba, Takuya and Sano, Shotaro and Yanase, Toshihiko and Ohta, Takeru and Koyama, Masanori},
	booktitle = {Proceedings of the 25th {ACM} {SIGKDD} International Conference on Knowledge Discovery and Data Mining},
	title = {Optuna: A Next-generation Hyperparameter Optimization Framework},
	year = {2019}}

@inproceedings{Bergstra_2011,
	address = {Red Hook, NY, USA},
	author = {Bergstra, James and Bardenet, R\'{e}mi and Bengio, Yoshua and K\'{e}gl, Bal\'{a}zs},
	booktitle = {Proceedings of the 25th International Conference on Neural Information Processing Systems},
	isbn = {9781618395993},
	location = {Granada, Spain},
	numpages = {9},
	pages = {2546--2554},
	publisher = {Curran Associates Inc.},
	series = {NIPS'11},
	title = {Algorithms for hyper-parameter optimization},
	year = {2011}}

@article{Balaguera_2012,
	author = {Balaguera-Antol{\'\i}nez, A. and S{\'a}nchez, Ariel G. and B{\"o}hringer, H. and Collins, C.},
	doi = {10.1111/j.1365-2966.2012.21685.x},
	issn = {0035-8711},
	journal = {MNRAS},
	month = {09},
	number = {3},
	pages = {2244-2254},
	title = {Constructing mock catalogues for the REFLEX II galaxy cluster sample},
	volume = {425},
	year = {2012}}

@inproceedings{Bergstra_2013,
	address = {Atlanta, Georgia, USA},
	author = {Bergstra, James and Yamins, Daniel and Cox, David},
	booktitle = {Proceedings of the 30th International Conference on Machine Learning},
	editor = {Dasgupta, Sanjoy and McAllester, David},
	month = {17--19 Jun},
	number = {1},
	pages = {115--123},
	publisher = {PMLR},
	series = {Proceedings of Machine Learning Research},
	title = {Making a Science of Model Search: Hyperparameter Optimization in Hundreds of Dimensions for Vision Architectures},
	url = {https://proceedings.mlr.press/v28/bergstra13.html},
	volume = {28},
	year = {2013}}

@article{Kingma_2014,
	author = {Diederik P. Kingma and Jimmy Ba},
	journal = {CoRR},
	title = {Adam: A Method for Stochastic Optimization},
	url = {https://api.semanticscholar.org/CorpusID:6628106},
	volume = {abs/1412.6980},
	year = {2014}}

@article{Loshchilov_2017,
	author = {Ilya Loshchilov and Frank Hutter},
	journal = {ArXiv},
	title = {Fixing Weight Decay Regularization in Adam},
	url = {https://api.semanticscholar.org/CorpusID:3312944},
	volume = {abs/1711.05101},
	year = {2017}}

@article{Watanabe2023,
	author = {{Watanabe}, Shuhei},
	doi = {10.48550/arXiv.2304.11127},
	journal = {arXiv e-prints},
	month = apr,
	title = {{Tree-Structured Parzen Estimator: Understanding Its Algorithm Components and Their Roles for Better Empirical Performance}},
	year = 2023}

@article{erosita_2024,
	author = {{Bulbul}, E. and {Liu}, A. and {Kluge}, M. and {Zhang}, X. and {Sanders}, J.~S. and {Bahar}, Y.~E. and {Ghirardini}, V. and {Artis}, E. and {Seppi}, R. and {Garrel}, C. and {Ramos-Ceja}, M.~E. and {Comparat}, J. and {Balzer}, F. and {B{\"o}ckmann}, K. and {Br{\"u}ggen}, M. and {Clerc}, N. and {Dennerl}, K. and {Dolag}, K. and {Freyberg}, M. and {Grandis}, S. and {Gruen}, D. and {Kleinebreil}, F. and {Krippendorf}, S. and {Lamer}, G. and {Merloni}, A. and {Migkas}, K. and {Nandra}, K. and {Pacaud}, F. and {Predehl}, P. and {Reiprich}, T.~H. and {Schrabback}, T. and {Veronica}, A. and {Weller}, J. and {Zelmer}, S.},
	doi = {10.1051/0004-6361/202348264},
	eid = {A106},
	journal = {A\&A},
	month = may,
	pages = {A106},
	title = {{The SRG/eROSITA All-Sky Survey. The first catalog of galaxy clusters and groups in the Western Galactic Hemisphere}},
	volume = {685},
	year = 2024}

@article{Villanueva_Domingo_2022,
	author = {Villanueva-Domingo, Pablo and Villaescusa-Navarro, Francisco},
	doi = {10.3847/1538-4357/ac8930},
	journal = {ApJ},
	month = oct,
	number = {2},
	pages = {115},
	publisher = {American Astronomical Society},
	title = {Learning Cosmology and Clustering with Cosmic Graphs},
	volume = {937},
	year = {2022}}

@article{Lee_2025,
	author = {{Lee}, Jun-Young and {Villaescusa-Navarro}, Francisco},
	doi = {10.3847/1538-4357/ade806},
	eid = {47},
	journal = {ApJ},
	month = aug,
	number = {1},
	pages = {47},
	title = {{Cosmology with Topological Deep Learning}},
	volume = {989},
	year = 2025}

@inproceedings{Balla_2024,
	author = {Julia Balla and Siddharth Mishra-Sharma and Carolina Cuesta-Lazaro and Tommi Jaakkola and Tess Smidt},
	booktitle = {The Third Learning on Graphs Conference},
	title = {A Cosmic-Scale Benchmark for Symmetry-Preserving Data Processing},
	url = {https://openreview.net/forum?id=t8yFkSAsLq},
	year = {2024}}

@article{Battaglia_2018,
	author = {{Battaglia}, Peter W. and {Hamrick}, Jessica B. and {Bapst}, Victor and {Sanchez-Gonzalez}, Alvaro and {Zambaldi}, Vinicius and {Malinowski}, Mateusz and {Tacchetti}, Andrea and {Raposo}, David and {Santoro}, Adam and {Faulkner}, Ryan and {Gulcehre}, Caglar and {Song}, Francis and {Ballard}, Andrew and {Gilmer}, Justin and {Dahl}, George and {Vaswani}, Ashish and {Allen}, Kelsey and {Nash}, Charles and {Langston}, Victoria and {Dyer}, Chris and {Heess}, Nicolas and {Wierstra}, Daan and {Kohli}, Pushmeet and {Botvinick}, Matt and {Vinyals}, Oriol and {Li}, Yujia and {Pascanu}, Razvan},
	doi = {10.48550/arXiv.1806.01261},
	journal = {arXiv e-prints},
	month = jun,
	title = {{Relational inductive biases, deep learning, and graph networks}},
	year = 2018}

@article{de_Santi_2023,
	author = {de Santi, Natal{\'\i} S. M. and Shao, Helen and Villaescusa-Navarro, Francisco and Abramo, L. Raul and Teyssier, Romain and Villanueva-Domingo, Pablo and Ni, Yueying and Angl{\'e}s-Alc{\'a}zar, Daniel and Genel, Shy and Hern{\'a}ndez-Mart{\'\i}nez, Elena and Steinwandel, Ulrich P. and Lovell, Christopher C. and Dolag, Klaus and Castro, Tiago and Vogelsberger, Mark},
	doi = {10.3847/1538-4357/acd1e2},
	journal = {The Astrophysical Journal},
	month = {jul},
	number = {1},
	pages = {69},
	publisher = {The American Astronomical Society},
	title = {Robust Field-level Likelihood-free Inference with Galaxies},
	volume = {952},
	year = {2023}}

@article{Makinen_2022,
	author = {{Makinen}, T. Lucas and {Charnock}, Tom and {Lemos}, Pablo and {Porqueres}, Natalia and {Heavens}, Alan F. and {Wandelt}, Benjamin D.},
	doi = {10.21105/astro.2207.05202},
	eid = {18},
	journal = {The Open Journal of Astrophysics},
	month = dec,
	number = {1},
	pages = {18},
	title = {{The Cosmic Graph: Optimal Information Extraction from Large-Scale Structure using Catalogues}},
	volume = {5},
	year = 2022}

@article{Shao_2023,
	author = {{Shao}, Helen and {Villaescusa-Navarro}, Francisco and {Villanueva-Domingo}, Pablo and {Teyssier}, Romain and {Garrison}, Lehman H. and {Gatti}, Marco and {Inman}, Derek and {Ni}, Yueying and {Steinwandel}, Ulrich P. and {Kulkarni}, Mihir and {Visbal}, Eli and {Bryan}, Greg L. and {Angl{\'e}s-Alc{\'a}zar}, Daniel and {Castro}, Tiago and {Hern{\'a}ndez-Mart{\'\i}nez}, Elena and {Dolag}, Klaus},
	doi = {10.3847/1538-4357/acac7a},
	journal = {ApJ},
	month = feb,
	number = {1},
	pages = {27},
	title = {{Robust Field-level Inference of Cosmological Parameters with Dark Matter Halos}},
	volume = {944},
	year = 2023}

@book{Tsallis,
	author = {{Tsallis}, Constantino},
	doi = {10.1007/978-0-387-85359-8},
	publisher = {Springer New York},
	title = {{Introduction to Nonextensive Statistical Mechanics}},
	year = 2009}

@article{Stanek2010,
	author = {{Stanek}, R. and {Rasia}, E. and {Evrard}, A.~E. and {Pearce}, F. and {Gazzola}, L.},
	doi = {10.1088/0004-637X/715/2/1508},
	journal = {ApJ},
	month = jun,
	number = {2},
	pages = {1508-1523},
	title = {{Massive Halos in Millennium Gas Simulations: Multivariate Scaling Relations}},
	volume = {715},
	year = 2010}

@article{Reiprich2006,
	author = {{Reiprich}, T.~H.},
	doi = {10.1051/0004-6361:20065525},
	journal = {A\&A},
	month = jul,
	number = {3},
	pages = {L39-L42},
	title = {{The galaxy cluster X-ray luminosity-gravitational mass relation in the light of the WMAP 3rd year data}},
	volume = {453},
	year = 2006}

@article{Bairagi_2025,
	author = {{Bairagi}, Anirban and {Wandelt}, Benjamin and {Villaescusa-Navarro}, Francisco},
	doi = {10.48550/arXiv.2503.13755},
	journal = {arXiv e-prints},
	month = mar,
	title = {{How many simulations do we need for simulation-based inference in cosmology?}},
	year = 2025}

@article{Kacprzak_2023,
	author = {Kacprzak, Tomasz and Fluri, Janis and Schneider, Aurel and Refregier, Alexandre and Stadel, Joachim},
	doi = {10.1088/1475-7516/2023/02/050},
	journal = {JCAP},
	month = {feb},
	number = {02},
	pages = {050},
	publisher = {IOP Publishing},
	title = {CosmoGridV1: a simulated LCDM theory prediction for map-level cosmological inference},
	volume = {2023},
	year = {2023}}

@article{Chen_2025,
	author = {Chen, Zhao and Yu, Yu and Han, Jiaxin and Jing, Yipeng},
	day = {06},
	doi = {10.1007/s11433-025-2671-0},
	issn = {1869-1927},
	journal = {Science China Physics, Mechanics {\&} Astronomy},
	month = {Jun},
	number = {8},
	pages = {289512},
	title = {CSST cosmological emulator I: Matter power spectrum emulation with one percent accuracy to k = 10h Mpc-1},
	volume = {68},
	year = {2025}}

@article{DeRose_2023,
	author = {DeRose, Joseph and Kokron, Nickolas and Banerjee, Arka and Chen, Shi-Fan and White, Martin and Wechsler, Risa and Storey-Fisher, Kate and Tinker, Jeremy and Zhai, Zhongxu},
	doi = {10.1088/1475-7516/2023/07/054},
	journal = {JCAP},
	month = {jul},
	number = {07},
	pages = {054},
	publisher = {IOP Publishing},
	title = {Aemulus ν: precise predictions for matter and biased tracer power spectra in the presence of neutrinos},
	volume = {2023},
	year = {2023}}

@article{Sobol_1967,
	author = {I.M Sobol'},
	doi = {https://doi.org/10.1016/0041-5553(67)90144-9},
	issn = {0041-5553},
	journal = {USSR Computational Mathematics and Mathematical Physics},
	number = {4},
	pages = {86-112},
	title = {On the distribution of points in a cube and the approximate evaluation of integrals},
	volume = {7},
	year = {1967}}

@article{Planck2018,
	author = {{Planck Collaboration} and {Aghanim}, N. and {Akrami}, Y. and {Ashdown}, M. and {Aumont}, J. and {Baccigalupi}, C. and {Ballardini}, M. and {Banday}, A.~J. and {Barreiro}, R.~B. and {Bartolo}, N. and {Basak}, S. and {Battye}, R. and {Benabed}, K. and {Bernard}, J. -P. and {Bersanelli}, M. and {Bielewicz}, P. and {Bock}, J.~J. and {Bond}, J.~R. and {Borrill}, J. and {Bouchet}, F.~R. and {Boulanger}, F. and {Bucher}, M. and {Burigana}, C. and {Butler}, R.~C. and {Calabrese}, E. and {Cardoso}, J. -F. and {Carron}, J. and {Challinor}, A. and {Chiang}, H.~C. and {Chluba}, J. and {Colombo}, L.~P.~L. and {Combet}, C. and {Contreras}, D. and {Crill}, B.~P. and {Cuttaia}, F. and {de Bernardis}, P. and {de Zotti}, G. and {Delabrouille}, J. and {Delouis}, J. -M. and {Di Valentino}, E. and {Diego}, J.~M. and {Dor{\'e}}, O. and {Douspis}, M. and {Ducout}, A. and {Dupac}, X. and {Dusini}, S. and {Efstathiou}, G. and {Elsner}, F. and {En{\ss}lin}, T.~A. and {Eriksen}, H.~K. and {Fantaye}, Y. and {Farhang}, M. and {Fergusson}, J. and {Fernandez-Cobos}, R. and {Finelli}, F. and {Forastieri}, F. and {Frailis}, M. and {Fraisse}, A.~A. and {Franceschi}, E. and {Frolov}, A. and {Galeotta}, S. and {Galli}, S. and {Ganga}, K. and {G{\'e}nova-Santos}, R.~T. and {Gerbino}, M. and {Ghosh}, T. and {Gonz{\'a}lez-Nuevo}, J. and {G{\'o}rski}, K.~M. and {Gratton}, S. and {Gruppuso}, A. and {Gudmundsson}, J.~E. and {Hamann}, J. and {Handley}, W. and {Hansen}, F.~K. and {Herranz}, D. and {Hildebrandt}, S.~R. and {Hivon}, E. and {Huang}, Z. and {Jaffe}, A.~H. and {Jones}, W.~C. and {Karakci}, A. and {Keih{\"a}nen}, E. and {Keskitalo}, R. and {Kiiveri}, K. and {Kim}, J. and {Kisner}, T.~S. and {Knox}, L. and {Krachmalnicoff}, N. and {Kunz}, M. and {Kurki-Suonio}, H. and {Lagache}, G. and {Lamarre}, J. -M. and {Lasenby}, A. and {Lattanzi}, M. and {Lawrence}, C.~R. and {Le Jeune}, M. and {Lemos}, P. and {Lesgourgues}, J. and {Levrier}, F. and {Lewis}, A. and {Liguori}, M. and {Lilje}, P.~B. and {Lilley}, M. and {Lindholm}, V. and {L{\'o}pez-Caniego}, M. and {Lubin}, P.~M. and {Ma}, Y. -Z. and {Mac{\'\i}as-P{\'e}rez}, J.~F. and {Maggio}, G. and {Maino}, D. and {Mandolesi}, N. and {Mangilli}, A. and {Marcos-Caballero}, A. and {Maris}, M. and {Martin}, P.~G. and {Martinelli}, M. and {Mart{\'\i}nez-Gonz{\'a}lez}, E. and {Matarrese}, S. and {Mauri}, N. and {McEwen}, J.~D. and {Meinhold}, P.~R. and {Melchiorri}, A. and {Mennella}, A. and {Migliaccio}, M. and {Millea}, M. and {Mitra}, S. and {Miville-Desch{\^e}nes}, M. -A. and {Molinari}, D. and {Montier}, L. and {Morgante}, G. and {Moss}, A. and {Natoli}, P. and {N{\o}rgaard-Nielsen}, H.~U. and {Pagano}, L. and {Paoletti}, D. and {Partridge}, B. and {Patanchon}, G. and {Peiris}, H.~V. and {Perrotta}, F. and {Pettorino}, V. and {Piacentini}, F. and {Polastri}, L. and {Polenta}, G. and {Puget}, J. -L. and {Rachen}, J.~P. and {Reinecke}, M. and {Remazeilles}, M. and {Renzi}, A. and {Rocha}, G. and {Rosset}, C. and {Roudier}, G. and {Rubi{\~n}o-Mart{\'\i}n}, J.~A. and {Ruiz-Granados}, B. and {Salvati}, L. and {Sandri}, M. and {Savelainen}, M. and {Scott}, D. and {Shellard}, E.~P.~S. and {Sirignano}, C. and {Sirri}, G. and {Spencer}, L.~D. and {Sunyaev}, R. and {Suur-Uski}, A. -S. and {Tauber}, J.~A. and {Tavagnacco}, D. and {Tenti}, M. and {Toffolatti}, L. and {Tomasi}, M. and {Trombetti}, T. and {Valenziano}, L. and {Valiviita}, J. and {Van Tent}, B. and {Vibert}, L. and {Vielva}, P. and {Villa}, F. and {Vittorio}, N. and {Wandelt}, B.~D. and {Wehus}, I.~K. and {White}, M. and {White}, S.~D.~M. and {Zacchei}, A. and {Zonca}, A.},
	doi = {10.1051/0004-6361/201833910},
	eid = {A6},
	journal = {A\&A},
	month = sep,
	pages = {A6},
	title = {{Planck 2018 results. VI. Cosmological parameters}},
	volume = {641},
	year = 2020}

@inproceedings{pytorch,
	author = {Ansel, Jason and Yang, Edward and He, Horace and Gimelshein, Natalia and Jain, Animesh and Voznesensky, Michael and Bao, Bin and Bell, Peter and Berard, David and Burovski, Evgeni and Chauhan, Geeta and Chourdia, Anjali and Constable, Will and Desmaison, Alban and DeVito, Zachary and Ellison, Elias and Feng, Will and Gong, Jiong and Gschwind, Michael and Hirsh, Brian and Huang, Sherlock and Kalambarkar, Kshiteej and Kirsch, Laurent and Lazos, Michael and Lezcano, Mario and Liang, Yanbo and Liang, Jason and Lu, Yinghai and Luk, CK and Maher, Bert and Pan, Yunjie and Puhrsch, Christian and Reso, Matthias and Saroufim, Mark and Siraichi, Marcos Yukio and Suk, Helen and Suo, Michael and Tillet, Phil and Wang, Eikan and Wang, Xiaodong and Wen, William and Zhang, Shunting and Zhao, Xu and Zhou, Keren and Zou, Richard and Mathews, Ajit and Chanan, Gregory and Wu, Peng and Chintala, Soumith},
	booktitle = {29th ACM International Conference on Architectural Support for Programming Languages and Operating Systems, Volume 2 (ASPLOS '24)},
	doi = {10.1145/3620665.3640366},
	month = apr,
	publisher = {ACM},
	title = {{PyTorch 2: Faster Machine Learning Through Dynamic Python Bytecode Transformation and Graph Compilation}},
	year = {2024}}

@article{Balaguera-Antolinez_2014,
	author = {{Balaguera-Antol{\'\i}nez}, Andr{\'e}s},
	doi = {10.1051/0004-6361/201322029},
	journal = {A\&A},
	pages = {A141},
	title = {What can the spatial distribution of galaxy clusters tell about their scaling relations?},
	volume = 563,
	year = 2014}

@article{Sefusatti_2016,
	author = {Sefusatti, E. and Crocce, M. and Scoccimarro, R. and Couchman, H. M. P.},
	doi = {10.1093/mnras/stw1229},
	issn = {0035-8711},
	journal = {MNRAS},
	month = {05},
	number = {4},
	pages = {3624-3636},
	title = {Accurate estimators of correlation functions in Fourier space},
	volume = {460},
	year = {2016}}

@article{Fumagalli_2025,
	author = {{Fumagalli}, A. and {Costanzi}, M. and {Castro}, T. and {Saro}, A. and {Borgani}, S. and {Romanello}, M. and {Marulli}, F. and {Tsaprazi}, E. and {Monaco}, P. and {Altieri}, B. and {Amara}, A. and {Amendola}, L. and {Andreon}, S. and {Auricchio}, N. and {Baccigalupi}, C. and {Baldi}, M. and {Balestra}, A. and {Bardelli}, S. and {Biviano}, A. and {Branchini}, E. and {Brescia}, M. and {Camera}, S. and {Ca{\~n}as-Herrera}, G. and {Capobianco}, V. and {Carbone}, C. and {Carretero}, J. and {Casas}, S. and {Castellano}, M. and {Castignani}, G. and {Cavuoti}, S. and {Chambers}, K.~C. and {Cimatti}, A. and {Colodro-Conde}, C. and {Congedo}, G. and {Conversi}, L. and {Copin}, Y. and {Courbin}, F. and {Courtois}, H.~M. and {Da Silva}, A. and {Degaudenzi}, H. and {de la Torre}, S. and {De Lucia}, G. and {Di Giorgio}, A.~M. and {Dole}, H. and {Douspis}, M. and {Dubath}, F. and {Duncan}, C.~A.~J. and {Dupac}, X. and {Dusini}, S. and {Escoffier}, S. and {Farina}, M. and {Farinelli}, R. and {Faustini}, F. and {Ferriol}, S. and {Finelli}, F. and {Fosalba}, P. and {Fourmanoit}, N. and {Frailis}, M. and {Franceschi}, E. and {Fumana}, M. and {Galeotta}, S. and {George}, K. and {Gillis}, B. and {Giocoli}, C. and {Gracia-Carpio}, J. and {Grazian}, A. and {Grupp}, F. and {Guzzo}, L. and {Haugan}, S.~V.~H. and {Holmes}, W. and {Hormuth}, F. and {Hornstrup}, A. and {Jahnke}, K. and {Jhabvala}, M. and {Joachimi}, B. and {Keih{\"a}nen}, E. and {Kermiche}, S. and {Kiessling}, A. and {Kubik}, B. and {K{\"u}mmel}, M. and {Kunz}, M. and {Kurki-Suonio}, H. and {Le Brun}, A.~M.~C. and {Ligori}, S. and {Lilje}, P.~B. and {Lindholm}, V. and {Lloro}, I. and {Mainetti}, G. and {Maino}, D. and {Maiorano}, E. and {Mansutti}, O. and {Marggraf}, O. and {Martinelli}, M. and {Martinet}, N. and {Massey}, R.~J. and {Medinaceli}, E. and {Mei}, S. and {Mellier}, Y. and {Meneghetti}, M. and {Merlin}, E. and {Meylan}, G. and {Mohr}, J.~J. and {Mora}, A. and {Moresco}, M. and {Moscardini}, L. and {Munari}, E. and {Nakajima}, R. and {Neissner}, C. and {Niemi}, S. -M. and {Padilla}, C. and {Paltani}, S. and {Pasian}, F. and {Pedersen}, K. and {Pettorino}, V. and {Pires}, S. and {Polenta}, G. and {Poncet}, M. and {Popa}, L.~A. and {Pozzetti}, L. and {Raison}, F. and {Rebolo}, R. and {Renzi}, A. and {Rhodes}, J. and {Riccio}, G. and {Romelli}, E. and {Roncarelli}, M. and {Rosset}, C. and {Saglia}, R. and {Sakr}, Z. and {S{\'a}nchez}, A.~G. and {Sapone}, D. and {Sartoris}, B. and {Schneider}, P. and {Schrabback}, T. and {Secroun}, A. and {Sefusatti}, E. and {Seidel}, G. and {Seiffert}, M. and {Serrano}, S. and {Simon}, P. and {Sirignano}, C. and {Sirri}, G. and {Spurio Mancini}, A. and {Stanco}, L. and {Steinwagner}, J. and {Tallada-Cresp{\'\i}}, P. and {Tavagnacco}, D. and {Taylor}, A.~N. and {Tereno}, I. and {Tessore}, N. and {Toft}, S. and {Toledo-Moreo}, R. and {Torradeflot}, F. and {Tutusaus}, I. and {Valenziano}, L. and {Valiviita}, J. and {Vassallo}, T. and {Verdoes Kleijn}, G. and {Veropalumbo}, A. and {Wang}, Y. and {Weller}, J. and {Zamorani}, G. and {Zerbi}, F.~M. and {Zucca}, E. and {Burigana}, C. and {Gabarra}, L. and {Maturi}, M. and {Porciani}, C. and {Scottez}, V. and {Sereno}, M. and {Viel}, M.},
	doi = {10.48550/arXiv.2510.13509},
	journal = {arXiv e-prints},
	month = oct,
	title = {{Euclid: Exploring observational systematics in cluster cosmology -- a comprehensive analysis of cluster counts and clustering}},
	year = 2025}

@article{Kaiser_1987,
	adsurl = {https://ui.adsabs.harvard.edu/abs/1987MNRAS.227....1K},
	author = {{Kaiser}, Nick},
	doi = {10.1093/mnras/227.1.1},
	journal = {MNRAS},
	month = jul,
	pages = {1-21},
	title = {{Clustering in real space and in redshift space}},
	volume = {227},
	year = 1987}

@article{Alcock_1979,
	author = {Alcock, Charles and Paczy{\'{n}}ski, Bohdan},
	day = {01},
	doi = {10.1038/281358a0},
	issn = {1476-4687},
	journal = {Nature},
	month = {Oct},
	number = {5730},
	pages = {358-359},
	title = {An evolution free test for non-zero cosmological constant},
	volume = {281},
	year = {1979}}

@article{Marulli_2018,
	author = {Marulli, F. and {Veropalumbo, A.} and {Sereno, M.} and {Moscardini, L.} and {Pacaud, F.} and {Pierre, M.} and {Plionis, M.} and {Cappi, A.} and {Adami, C.} and {Alis, S.} and {Altieri, B.} and {Birkinshaw, M.} and {Ettori, S.} and {Faccioli, L.} and {Gastaldello, F.} and {Koulouridis, E.} and {Lidman, C.} and {Le F{\`e}vre, J.-P.} and {Maurogordato, S.} and {Poggianti, B.} and {Pompei, E.} and {Sadibekova, T.} and {Valtchanov, I.}},
	doi = {10.1051/0004-6361/201833238},
	journal = {A\&A},
	pages = {A1},
	title = {The XXL Survey - XVI. The clustering of X-ray selected galaxy clusters at z ~ 0.3},
	volume = 620,
	year = 2018}

@article{Min_2024,
	author = {Min, Zhiwei and Xiao, Xu and Ding, Jiacheng and Xiao, Liang and Jiang, Jie and Wu, Donglin and Lin, Qiufan and Wang, Yang and Liu, Shuai and Chen, Zhixin and Li, Xiangru and Zhang, Jinqu and Zhang, Le and Li, Xiao-Dong},
	doi = {10.1103/PhysRevD.110.063531},
	issue = {6},
	journal = {Phys. Rev. D},
	month = {Sep},
	numpages = {17},
	pages = {063531},
	publisher = {American Physical Society},
	title = {Deep learning for cosmological parameter inference from a dark matter halo density field},
	volume = {110},
	year = {2024}}

@article{Villaescusa-Navarro_2022,
	author = {Villaescusa-Navarro, Francisco and Genel, Shy and Angl{\'e}s-Alc{\'a}zar, Daniel and Thiele, Leander and Dave, Romeel and Narayanan, Desika and Nicola, Andrina and Li, Yin and Villanueva-Domingo, Pablo and Wandelt, Benjamin and Spergel, David N. and Somerville, Rachel S. and Zorrilla Matilla, Jose Manuel and Mohammad, Faizan G. and Hassan, Sultan and Shao, Helen and Wadekar, Digvijay and Eickenberg, Michael and Wong, Kaze W. K. and Contardo, Gabriella and Jo, Yongseok and Moser, Emily and Lau, Erwin T. and Machado Poletti Valle, Luis Fernando and Perez, Lucia A. and Nagai, Daisuke and Battaglia, Nicholas and Vogelsberger, Mark},
	doi = {10.3847/1538-4365/ac5ab0},
	journal = {The Astrophysical Journal Supplement Series},
	month = {apr},
	number = {2},
	pages = {61},
	publisher = {The American Astronomical Society},
	title = {The CAMELS Multifield Data Set: Learning the Universe's Fundamental Parameters with Artificial Intelligence},
	volume = {259},
	year = {2022}}

@article{Lemos_2024,
	author = {Lemos, Pablo and Parker, Liam and Hahn, ChangHoon and Ho, Shirley and Eickenberg, Michael and Hou, Jiamin and Massara, Elena and Modi, Chirag and Dizgah, Azadeh Moradinezhad and Blancard, Bruno R\'egaldo-Saint and Spergel, David},
	collaboration = {SimBIG Collaboration},
	doi = {10.1103/PhysRevD.109.083536},
	issue = {8},
	journal = {Phys. Rev. D},
	month = {Apr},
	numpages = {12},
	pages = {083536},
	publisher = {American Physical Society},
	title = {Field-level simulation-based inference of galaxy clustering with convolutional neural networks},
	volume = {109},
	year = {2024}}

@article{Sharma_2024,
	author = {Sharma, Divij and Dai, Biwei and Seljak, Uro{\v s}},
	doi = {10.1088/1475-7516/2024/08/010},
	journal = {JCAP},
	month = {aug},
	number = {08},
	pages = {010},
	publisher = {IOP Publishing},
	title = {A comparative study of cosmological constraints from weak lensing using Convolutional Neural Networks},
	volume = {2024},
	year = {2024}}

@article{Gupta_2018,
	author = {Gupta, Arushi and Matilla, Jos\'e Manuel Zorrilla and Hsu, Daniel and Haiman, Zolt\'an},
	doi = {10.1103/PhysRevD.97.103515},
	issue = {10},
	journal = {Phys. Rev. D},
	numpages = {15},
	pages = {103515},
	publisher = {American Physical Society},
	title = {Non-Gaussian information from weak lensing data via deep learning},
	volume = {97},
	year = {2018}}

@inproceedings{Ravanbakhsh_2016,
	address = {New York, New York, USA},
	author = {Ravanbakhsh, Siamak and Oliva, Junier and Fromenteau, Sebastian and Price, Layne and Ho, Shirley and Schneider, Jeff and Poczos, Barnabas},
	booktitle = {Proceedings of The 33rd International Conference on Machine Learning},
	month = {20--22 Jun},
	pages = {2407--2416},
	publisher = {PMLR},
	series = {Proceedings of Machine Learning Research},
	title = {Estimating Cosmological Parameters from the Dark Matter Distribution},
	volume = {48},
	year = {2016}}

@article{LeCun_1989,
	author = {LeCun, Y. and Boser, B. and Denker, J. S. and Henderson, D. and Howard, R. E. and Hubbard, W. and Jackel, L. D.},
	doi = {10.1162/neco.1989.1.4.541},
	journal = {Neural Computation},
	number = {4},
	pages = {541-551},
	title = {Backpropagation Applied to Handwritten Zip Code Recognition},
	volume = {1},
	year = {1989}}

@article{Jeffrey_2025,
	author = {{Jeffrey}, N. and {Whiteway}, L. and {Gatti}, M. and {Williamson}, J. and {Alsing}, J. and {Porredon}, A. and {Prat}, J. and {Doux}, C. and {Jain}, B. and {Chang}, C. and {Cheng}, T.-Y. and {Kacprzak}, T. and {Lemos}, P. and {Alarcon}, A. and {Amon}, A. and {Bechtol}, K. and {Becker}, M.~R. and {Bernstein}, G.~M. and {Campos}, A. and {Carnero Rosell}, A. and {Chen}, R. and {Choi}, A. and {DeRose}, J. and {Drlica-Wagner}, A. and {Eckert}, K. and {Everett}, S. and {Fert{\'e}}, A. and {Gruen}, D. and {Gruendl}, R.~A. and {Herner}, K. and {Jarvis}, M. and {McCullough}, J. and {Myles}, J. and {Navarro-Alsina}, A. and {Pandey}, S. and {Raveri}, M. and {Rollins}, R.~P. and {Rykoff}, E.~S. and {S{\'a}nchez}, C. and {Secco}, L.~F. and {Sevilla-Noarbe}, I. and {Sheldon}, E. and {Shin}, T. and {Troxel}, M.~A. and {Tutusaus}, I. and {Varga}, T.~N. and {Yanny}, B. and {Yin}, B. and {Zuntz}, J. and {Aguena}, M. and {Allam}, S.~S. and {Alves}, O. and {Bacon}, D. and {Bocquet}, S. and {Brooks}, D. and {da Costa}, L.~N. and {Davis}, T.~M. and {De Vicente}, J. and {Desai}, S. and {Diehl}, H.~T. and {Ferrero}, I. and {Frieman}, J. and {Garc{\'\i}a-Bellido}, J. and {Gaztanaga}, E. and {Giannini}, G. and {Gutierrez}, G. and {Hinton}, S.~R. and {Hollowood}, D.~L. and {Honscheid}, K. and {Huterer}, D. and {James}, D.~J. and {Lahav}, O. and {Lee}, S. and {Marshall}, J.~L. and {Mena-Fern{\'a}ndez}, J. and {Miquel}, R. and {Pieres}, A. and {Plazas Malag{\'o}n}, A.~A. and {Roodman}, A. and {Sako}, M. and {Sanchez}, E. and {Sanchez Cid}, D. and {Smith}, M. and {Suchyta}, E. and {Swanson}, M.~E.~C. and {Tarle}, G. and {Tucker}, D.~L. and {Weaverdyck}, N. and {Weller}, J. and {Wiseman}, P. and {Yamamoto}, M.},
	doi = {10.1093/mnras/stae2629},
	journal = {MNRAS},
	month = jan,
	number = {2},
	pages = {1303-1322},
	title = {{Dark energy survey year 3 results: likelihood-free, simulation-based wCDM inference with neural compression of weak-lensing map statistics}},
	volume = {536},
	year = 2025}

@article{Jamieson_2023,
	author = {Jamieson, Drew and Li, Yin and de Oliveira, Renan Alves and Villaescusa-Navarro, Francisco and Ho, Shirley and Spergel, David N.},
	doi = {10.3847/1538-4357/acdb6c},
	journal = {ApJ},
	month = {jul},
	number = {2},
	pages = {145},
	publisher = {The American Astronomical Society},
	title = {Field-level Neural Network Emulator for Cosmological N-body Simulations},
	volume = {952},
	year = {2023}}

@article{He_2019,
	author = {{He}, Siyu and {Li}, Yin and {Feng}, Yu and {Ho}, Shirley and {Ravanbakhsh}, Siamak and {Chen}, Wei and {P{\'o}czos}, Barnab{\'a}s},
	doi = {10.1073/pnas.1821458116},
	journal = {Proceedings of the National Academy of Science},
	month = jul,
	number = {28},
	pages = {13825-13832},
	title = {{Learning to predict the cosmological structure formation}},
	volume = {116},
	year = 2019}

@article{deOliveira_2020,
	author = {{Alves de Oliveira}, Renan and {Li}, Yin and {Villaescusa-Navarro}, Francisco and {Ho}, Shirley and {Spergel}, David N.},
	doi = {10.48550/arXiv.2012.00240},
	journal = {arXiv e-prints},
	month = nov,
	title = {{Fast and Accurate Non-Linear Predictions of Universes with Deep Learning}},
	year = 2020}

@article{Jamieson_2025,
	author = {{Jamieson}, Drew and {Li}, Yin and {Villaescusa-Navarro}, Francisco and {Ho}, Shirley and {Spergel}, David N.},
	doi = {10.1088/1475-7516/2025/03/072},
	journal = {jcap},
	month = mar,
	number = {3},
	pages = {072},
	title = {{Field-level emulation of cosmic structure formation with cosmology and redshift dependence}},
	volume = {2025},
	year = 2025}

@article{Kaushal_2022,
	author = {Kaushal, Neerav and Villaescusa-Navarro, Francisco and Giusarma, Elena and Li, Yin and Hawry, Conner and Reyes, Mauricio},
	doi = {10.3847/1538-4357/ac5c4a},
	journal = {ApJ},
	month = {may},
	number = {2},
	pages = {115},
	publisher = {The American Astronomical Society},
	title = {NECOLA: Toward a Universal Field-level Cosmological Emulator},
	volume = {930},
	year = {2022}}

@article{Riess_1998,
	author = {Riess, Adam G. and Filippenko, Alexei V. and Challis, Peter and Clocchiatti, Alejandro and Diercks, Alan and Garnavich, Peter M. and Gilliland, Ron L. and Hogan, Craig J. and Jha, Saurabh and Kirshner, Robert P. and Leibundgut, B. and Phillips, M. M. and Reiss, David and Schmidt, Brian P. and Schommer, Robert A. and Smith, R. Chris and Spyromilio, J. and Stubbs, Christopher and Suntzeff, Nicholas B. and Tonry, John},
	doi = {10.1086/300499},
	journal = {The Astronomical Journal},
	month = {sep},
	number = {3},
	pages = {1009},
	title = {Observational Evidence from Supernovae for an Accelerating Universe and a Cosmological Constant},
	volume = {116},
	year = {1998}}

@article{Perlmutter_1999,
	author = {Perlmutter, S. and Aldering, G. and Goldhaber, G. and Knop, R. A. and Nugent, P. and Castro, P. G. and Deustua, S. and Fabbro, S. and Goobar, A. and Groom, D. E. and Hook, I. M. and Kim, A. G. and Kim, M. Y. and Lee, J. C. and Nunes, N. J. and Pain, R. and Pennypacker, C. R. and Quimby, R. and Lidman, C. and Ellis, R. S. and Irwin, M. and McMahon, R. G. and Ruiz-Lapuente, P. and Walton, N. and Schaefer, B. and Boyle, B. J. and Filippenko, A. V. and Matheson, T. and Fruchter, A. S. and Panagia, N. and Newberg, H. J. M. and Couch, W. J. and Project, The Supernova Cosmology},
	doi = {10.1086/307221},
	journal = {ApJ},
	month = {jun},
	number = {2},
	pages = {565},
	title = {Measurements of Ω and Λ from 42 High-Redshift Supernovae},
	volume = {517},
	year = {1999}}

@article{Mellier_2025,
	author = {{Euclid Collaboration} and {Mellier, Y.} and {Abdurro'uf} and {Acevedo Barroso, J. A.} and {Ach{\'u}carro, A.} and {Adamek, J.} and {Adam, R.} and {Addison, G. E.} and {Aghanim, N.} and {Aguena, M.} and {Ajani, V.} and {Akrami, Y.} and {Al-Bahlawan, A.} and {Alavi, A.} and {Albuquerque, I. S.} and {Alestas, G.} and {Alguero, G.} and {Allaoui, A.} and {Allen, S. W.} and {Allevato, V.} and {Alonso-Tetilla, A. V.} and {Altieri, B.} and {Alvarez-Candal, A.} and {Alvi, S.} and {Amara, A.} and {Amendola, L.} and {Amiaux, J.} and {Andika, I. T.} and {Andreon, S.} and {Andrews, A.} and {Angora, G.} and {Angulo, R. E.} and {Annibali, F.} and {Anselmi, A.} and {Anselmi, S.} and {Arcari, S.} and {Archidiacono, M.} and {Aric{\`o}, G.} and {Arnaud, M.} and {Arnouts, S.} and {Asgari, M.} and {Asorey, J.} and {Atayde, L.} and {Atek, H.} and {Atrio-Barandela, F.} and {Aubert, M.} and {Aubourg, E.} and {Auphan, T.} and {Auricchio, N.} and {Aussel, B.} and {Aussel, H.} and {Avelino, P. P.} and {Avgoustidis, A.} and {Avila, S.} and {Awan, S.} and {Azzollini, R.} and {Baccigalupi, C.} and {Bachelet, E.} and {Bacon, D.} and {Baes, M.} and {Bagley, M. B.} and {Bahr-Kalus, B.} and {Balaguera-Antolinez, A.} and {Balbinot, E.} and {Balcells, M.} and {Baldi, M.} and {Baldry, I.} and {Balestra, A.} and {Ballardini, M.} and {Ballester, O.} and {Balogh, M.} and {Ba{\~n}ados, E.} and {Barbier, R.} and {Bardelli, S.} and {Baron, M.} and {Barreiro, T.} and {Barrena, R.} and {Barriere, J.-C.} and {Barros, B. J.} and {Barthelemy, A.} and {Bartolo, N.} and {Basset, A.} and {Battaglia, P.} and {Battisti, A. J.} and {Baugh, C. M.} and {Baumont, L.} and {Bazzanini, L.} and {Beaulieu, J.-P.} and {Beckmann, V.} and {Belikov, A. N.} and {Bel, J.} and {Bellagamba, F.} and {Bella, M.} and {Bellini, E.} and {Benabed, K.} and {Bender, R.} and {Benevento, G.} and {Bennett, C. L.} and {Benson, K.} and {Bergamini, P.} and {Bermejo-Climent, J. R.} and {Bernardeau, F.} and {Bertacca, D.} and {Berthe, M.} and {Berthier, J.} and {Bethermin, M.} and {Beutler, F.} and {Bevillon, C.} and {Bhargava, S.} and {Bhatawdekar, R.} and {Bianchi, D.} and {Bisigello, L.} and {Biviano, A.} and {Blake, R. P.} and {Blanchard, A.} and {Blazek, J.} and {Blot, L.} and {Bosco, A.} and {Bodendorf, C.} and {Boenke, T.} and {B{\"o}hringer, H.} and {Boldrini, P.} and {Bolzonella, M.} and {Bonchi, A.} and {Bonici, M.} and {Bonino, D.} and {Bonino, L.} and {Bonvin, C.} and {Bon, W.} and {Booth, J. T.} and {Borgani, S.} and {Borlaff, A. S.} and {Borsato, E.} and {Bosco, A.} and {Bose, B.} and {Botticella, M. T.} and {Boucaud, A.} and {Bouche, F.} and {Boucher, J. S.} and {Boutigny, D.} and {Bouvard, T.} and {Bouwens, R.} and {Bouy, H.} and {Bowler, R. A. A.} and {Bozza, V.} and {Bozzo, E.} and {Branchini, E.} and {Brando, G.} and {Brau-Nogue, S.} and {Brekke, P.} and {Bremer, M. N.} and {Brescia, M.} and {Breton, M.-A.} and {Brinchmann, J.} and {Brinckmann, T.} and {Brockley-Blatt, C.} and {Brodwin, M.} and {Brouard, L.} and {Brown, M. L.} and {Bruton, S.} and {Bucko, J.} and {Buddelmeijer, H.} and {Buenadicha, G.} and {Buitrago, F.} and {Burger, P.} and {Burigana, C.} and {Busillo, V.} and {Busonero, D.} and {Cabanac, R.} and {Cabayol-Garcia, L.} and {Cagliari, M. S.} and {Caillat, A.} and {Caillat, L.} and {Calabrese, M.} and {Calabro, A.} and {Calderone, G.} and {Calura, F.} and {Camacho Quevedo, B.} and {Camera, S.} and {Campos, L.} and {Ca{\~n}as-Herrera, G.} and {Candini, G. P.} and {Cantiello, M.} and {Capobianco, V.} and {Cappellaro, E.} and {Cappelluti, N.} and {Cappi, A.} and {Caputi, K. I.} and {Cara, C.} and {Carbone, C.} and {Cardone, V. F.} and {Carella, E.} and {Carlberg, R. G.} and {Carle, M.} and {Carminati, L.} and {Caro, F.} and {Carrasco, J. M.} and {Carretero, J.} and {Carrilho, P.} and {Carron Duque, J.} and {Carry, B.} and {Carvalho, A.} and {Carvalho, C. S.} and {Casas, R.} and {Casas, S.} and {Casenove, P.} and {Casey, C. M.} and {Cassata, P.} and {Castander, F. J.} and {Castelao, D.} and {Castellano, M.} and {Castiblanco, L.} and {Castignani, G.} and {Castro, T.} and {Cavet, C.} and {Cavuoti, S.} and {Chabaud, P.-Y.} and {Chambers, K. C.} and {Charles, Y.} and {Charlot, S.} and {Chartab, N.} and {Chary, R.} and {Chaumeil, F.} and {Cho, H.} and {Chon, G.} and {Ciancetta, E.} and {Ciliegi, P.} and {Cimatti, A.} and {Cimino, M.} and {Cioni, M.-R. L.} and {Claydon, R.} and {Cleland, C.} and {Cl{\'e}ment, B.} and {Clements, D. L.} and {Clerc, N.} and {Clesse, S.} and {Codis, S.} and {Cogato, F.} and {Colbert, J.} and {Cole, R. E.} and {Coles, P.} and {Collett, T. E.} and {Collins, R. S.} and {Colodro-Conde, C.} and {Colombo, C.} and {Combes, F.} and {Conforti, V.} and {Congedo, G.} and {Conseil, S.} and {Conselice, C. J.} and {Contarini, S.} and {Contini, T.} and {Conversi, L.} and {Cooray, A. R.} and {Copin, Y.} and {Corasaniti, P.-S.} and {Corcho-Caballero, P.} and {Corcione, L.} and {Cordes, O.} and {Corpace, O.} and {Correnti, M.} and {Costanzi, M.} and {Costille, A.} and {Courbin, F.} and {Courcoult Mifsud, L.} and {Courtois, H. M.} and {Cousinou, M.-C.} and {Covone, G.} and {Cowell, T.} and {Cragg, C.} and {Cresci, G.} and {Cristiani, S.} and {Crocce, M.} and {Cropper, M.} and {Crouzet, P. E.} and {Csizi, B.} and {Cuby, J.-G.} and {Cucchetti, E.} and {Cucciati, O.} and {Cuillandre, J.-C.} and {Cunha, P. A. C.} and {Cuozzo, V.} and {Daddi, E.} and {D'Addona, M.} and {Dafonte, C.} and {Dagoneau, N.} and {Dalessandro, E.} and {Dalton, G. B.} and {D'Amico, G.} and {Dannerbauer, H.} and {Danto, P.} and {Das, I.} and {Da Silva, A.} and {da Silva, R.} and {d'Assignies Doumerg, W.} and {Daste, G.} and {Davies, J. E.} and {Davini, S.} and {Dayal, P.} and {de Boer, T.} and {Decarli, R.} and {De Caro, B.} and {Degaudenzi, H.} and {Degni, G.} and {de Jong, J. T. A.} and {de la Bella, L. F.} and {de la Torre, S.} and {Delhaise, F.} and {Delley, D.} and {Delucchi, G.} and {De Lucia, G.} and {Denniston, J.} and {De Paolis, F.} and {De Petris, M.} and {Derosa, A.} and {Desai, S.} and {Desjacques, V.} and {Despali, G.} and {Desprez, G.} and {De Vicente-Albendea, J.} and {Deville, Y.} and {Dias, J. D. F.} and {D{\'\i}az-S{\'a}nchez, A.} and {Diaz, J. J.} and {Di Domizio, S.} and {Diego, J. M.} and {Di Ferdinando, D.} and {Di Giorgio, A. M.} and {Dimauro, P.} and {Dinis, J.} and {Dolag, K.} and {Dolding, C.} and {Dole, H.} and {Dom{\'\i}nguez S{\'a}nchez, H.} and {Dor{\'e}, O.} and {Dournac, F.} and {Douspis, M.} and {Dreihahn, H.} and {Droge, B.} and {Dryer, B.} and {Dubath, F.} and {Duc, P.-A.} and {Ducret, F.} and {Duffy, C.} and {Dufresne, F.} and {Duncan, C. A. J.} and {Dupac, X.} and {Duret, V.} and {Durrer, R.} and {Durret, F.} and {Dusini, S.} and {Ealet, A.} and {Eggemeier, A.} and {Eisenhardt, P. R. M.} and {Elbaz, D.} and {Elkhashab, M. Y.} and {Ellien, A.} and {Endicott, J.} and {Enia, A.} and {Erben, T.} and {Escartin Vigo, J. A.} and {Escoffier, S.} and {Escudero Sanz, I.} and {Essert, J.} and {Ettori, S.} and {Ezziati, M.} and {Fabbian, G.} and {Fabricius, M.} and {Fang, Y.} and {Farina, A.} and {Farina, M.} and {Farinelli, R.} and {Farrens, S.} and {Faustini, F.} and {Feltre, A.} and {Ferguson, A. M. N.} and {Ferrando, P.} and {Ferrari, A. G.} and {Ferr{\'e}-Mateu, A.} and {Ferreira, P. G.} and {Ferreras, I.} and {Ferrero, I.} and {Ferriol, S.} and {Ferruit, P.} and {Filleul, D.} and {Finelli, F.} and {Finkelstein, S. L.} and {Finoguenov, A.} and {Fiorini, B.} and {Flentge, F.} and {Focardi, P.} and {Fonseca, J.} and {Fontana, A.} and {Fontanot, F.} and {Fornari, F.} and {Fosalba, P.} and {Fossati, M.} and {Fotopoulou, S.} and {Fouchez, D.} and {Fourmanoit, N.} and {Frailis, M.} and {Fraix-Burnet, D.} and {Franceschi, E.} and {Franco, A.} and {Franzetti, P.} and {Freihoefer, J.} and {Frenk, C. S.} and {Frittoli, G.} and {Frugier, P.-A.} and {Frusciante, N.} and {Fumagalli, A.} and {Fumagalli, M.} and {Fumana, M.} and {Fu, Y.} and {Gabarra, L.} and {Galeotta, S.} and {Galluccio, L.} and {Ganga, K.} and {Gao, H.} and {Garc{\'\i}a-Bellido, J.} and {Garcia, K.} and {Gardner, J. P.} and {Garilli, B.} and {Gaspar-Venancio, L.-M.} and {Gasparetto, T.} and {Gautard, V.} and {Gavazzi, R.} and {Gaztanaga, E.} and {Genolet, L.} and {Genova Santos, R.} and {Gentile, F.} and {George, K.} and {Gerbino, M.} and {Ghaffari, Z.} and {Giacomini, F.} and {Gianotti, F.} and {Gibb, G. P. S.} and {Gillard, W.} and {Gillis, B.} and {Ginolfi, M.} and {Giocoli, C.} and {Girardi, M.} and {Giri, S. K.} and {Goh, L. W. K.} and {G{\'o}mez-Alvarez, P.} and {Gonzalez-Perez, V.} and {Gonzalez, A. H.} and {Gonzalez, E. J.} and {Gonzalez, J. C.} and {Gouyou Beauchamps, S.} and {Gozaliasl, G.} and {Gracia-Carpio, J.} and {Grandis, S.} and {Granett, B. R.} and {Granvik, M.} and {Grazian, A.} and {Gregorio, A.} and {Grenet, C.} and {Grillo, C.} and {Grupp, F.} and {Gruppioni, C.} and {Gruppuso, A.} and {Guerbuez, C.} and {Guerrini, S.} and {Guidi, M.} and {Guillard, P.} and {Gutierrez, C. M.} and {Guttridge, P.} and {Guzzo, L.} and {Gwyn, S.} and {Haapala, J.} and {Haase, J.} and {Haddow, C. R.} and {Hailey, M.} and {Hall, A.} and {Hall, D.} and {Hamaus, N.} and {Haridasu, B. S.} and {Harnois-D{\'e}raps, J.} and {Harper, C.} and {Hartley, W. G.} and {Hasinger, G.} and {Hassani, F.} and {Hatch, N. A.} and {Haugan, S. V. H.} and {H{\"a}u{\ss}ler, B.} and {Heavens, A.} and {Heisenberg, L.} and {Helmi, A.} and {Helou, G.} and {Hemmati, S.} and {Henares, K.} and {Herent, O.} and {Hern{\'a}ndez-Monteagudo, C.} and {Heuberger, T.} and {Hewett, P. C.} and {Heydenreich, S.} and {Hildebrandt, H.} and {Hirschmann, M.} and {Hjorth, J.} and {Hoar, J.} and {Hoekstra, H.} and {Holland, A. D.} and {Holliman, M. S.} and {Holmes, W.} and {Hook, I.} and {Horeau, B.} and {Hormuth, F.} and {Hornstrup, A.} and {Hosseini, S.} and {Hu, D.} and {Hudelot, P.} and {Hudson, M. J.} and {Huertas-Company, M.} and {Huff, E. M.} and {Hughes, A. C. N.} and {Humphrey, A.} and {Hunt, L. K.} and {Huynh, D. D.} and {Ibata, R.} and {Ichikawa, K.} and {Iglesias-Groth, S.} and {Ilbert, O.} and {Ili{\'c}, S.} and {Ingoglia, L.} and {Iodice, E.} and {Israel, H.} and {Israelsson, U. E.} and {Izzo, L.} and {Jablonka, P.} and {Jackson, N.} and {Jacobson, J.} and {Jafariyazani, M.} and {Jahnke, K.} and {Jain, B.} and {Jansen, H.} and {Jarvis, M. J.} and {Jasche, J.} and {Jauzac, M.} and {Jeffrey, N.} and {Jhabvala, M.} and {Jimenez-Teja, Y.} and {Jimenez Mu{\~n}oz, A.} and {Joachimi, B.} and {Johansson, P. H.} and {Joudaki, S.} and {Jullo, E.} and {Kajava, J. J. E.} and {Kang, Y.} and {Kannawadi, A.} and {Kansal, V.} and {Karagiannis, D.} and {K{\"a}rcher, M.} and {Kashlinsky, A.} and {Kazandjian, M. V.} and {Keck, F.} and {Keih{\"a}nen, E.} and {Kerins, E.} and {Kermiche, S.} and {Khalil, A.} and {Kiessling, A.} and {Kiiveri, K.} and {Kilbinger, M.} and {Kim, J.} and {King, R.} and {Kirkpatrick, C. C.} and {Kitching, T.} and {Kluge, M.} and {Knabenhans, M.} and {Knapen, J. H.} and {Knebe, A.} and {Kneib, J.-P.} and {Kohley, R.} and {Koopmans, L. V. E.} and {Koskinen, H.} and {Koulouridis, E.} and {Kou, R.} and {Kov{\'a}cs, A.} and {Kova{\v c}i{\'c}, I.} and {Kowalczyk, A.} and {Koyama, K.} and {Kraljic, K.} and {Krause, O.} and {Kruk, S.} and {Kubik, B.} and {Kuchner, U.} and {Kuijken, K.} and {K{\"u}mmel, M.} and {Kunz, M.} and {Kurki-Suonio, H.} and {Lacasa, F.} and {Lacey, C. G.} and {La Franca, F.} and {Lagarde, N.} and {Lahav, O.} and {Laigle, C.} and {La Marca, A.} and {La Marle, O.} and {Lamine, B.} and {Lam, M. C.} and {Lan{\c c}on, A.} and {Landt, H.} and {Langer, M.} and {Lapi, A.} and {Larcheveque, C.} and {Larsen, S. S.} and {Lattanzi, M.} and {Laudisio, F.} and {Laugier, D.} and {Laureijs, R.} and {Laurent, V.} and {Lavaux, G.} and {Lawrenson, A.} and {Lazanu, A.} and {Lazeyras, T.} and {Le Boulc'h, Q.} and {Le Brun, A. M. C.} and {Le Brun, V.} and {Leclercq, F.} and {Lee, S.} and {Le Graet, J.} and {Legrand, L.} and {Leirvik, K. N.} and {Le Jeune, M.} and {Lembo, M.} and {Le Mignant, D.} and {Lepinzan, M. D.} and {Lepori, F.} and {Le Reun, A.} and {Leroy, G.} and {Lesci, G. F.} and {Lesgourgues, J.} and {Leuzzi, L.} and {Levi, M. E.} and {Liaudat, T. I.} and {Libet, G.} and {Liebing, P.} and {Ligori, S.} and {Lilje, P. B.} and {Lin, C.-C.} and {Linde, D.} and {Linder, E.} and {Lindholm, V.} and {Linke, L.} and {Li, S.-S.} and {Liu, S. J.} and {Lloro, I.} and {Lobo, F. S. N.} and {Lodieu, N.} and {Lombardi, M.} and {Lombriser, L.} and {Lonare, P.} and {Longo, G.} and {L{\'o}pez-Caniego, M.} and {Lopez Lopez, X.} and {Lorenzo Alvarez, J.} and {Loureiro, A.} and {Loveday, J.} and {Lusso, E.} and {Macias-Perez, J.} and {Maciaszek, T.} and {Maggio, G.} and {Magliocchetti, M.} and {Magnard, F.} and {Magnier, E. A.} and {Magro, A.} and {Mahler, G.} and {Mainetti, G.} and {Maino, D.} and {Maiorano, E.} and {Maiorano, E.} and {Malavasi, N.} and {Mamon, G. A.} and {Mancini, C.} and {Mandelbaum, R.} and {Manera, M.} and {Manj{\'o}n-Garc{\'\i}a, A.} and {Mannucci, F.} and {Mansutti, O.} and {Manteiga Outeiro, M.} and {Maoli, R.} and {Maraston, C.} and {Marcin, S.} and {Marcos-Arenal, P.} and {Margalef-Bentabol, B.} and {Marggraf, O.} and {Marinucci, D.} and {Marinucci, M.} and {Markovic, K.} and {Marleau, F. R.} and {Marpaud, J.} and {Martignac, J.} and {Mart{\'\i}n-Fleitas, J.} and {Martin-Moruno, P.} and {Martin, E. L.} and {Martinelli, M.} and {Martinet, N.} and {Martin, H.} and {Martins, C. J. A. P.} and {Marulli, F.} and {Massari, D.} and {Massey, R.} and {Masters, D. C.} and {Matarrese, S.} and {Matsuoka, Y.} and {Matthew, S.} and {Maughan, B. J.} and {Mauri, N.} and {Maurin, L.} and {Maurogordato, S.} and {McCarthy, K.} and {McConnachie, A. W.} and {McCracken, H. J.} and {McDonald, I.} and {McEwen, J. D.} and {McPartland, C. J. R.} and {Medinaceli, E.} and {Mehta, V.} and {Mei, S.} and {Melchior, M.} and {Melin, J.-B.} and {M{\'e}nard, B.} and {Mendes, J.} and {Mendez-Abreu, J.} and {Meneghetti, M.} and {Mercurio, A.} and {Merlin, E.} and {Metcalf, R. B.} and {Meylan, G.} and {Migliaccio, M.} and {Mignoli, M.} and {Miller, L.} and {Miluzio, M.} and {Milvang-Jensen, B.} and {Mimoso, J. P.} and {Miquel, R.} and {Miyatake, H.} and {Mobasher, B.} and {Mohr, J. J.} and {Monaco, P.} and {Mongui{\'o}, M.} and {Montoro, A.} and {Mora, A.} and {Moradinezhad Dizgah, A.} and {Moresco, M.} and {Moretti, C.} and {Morgante, G.} and {Morisset, N.} and {Moriya, T. J.} and {Morris, P. W.} and {Mortlock, D. J.} and {Moscardini, L.} and {Mota, D. F.} and {Mottet, S.} and {Moustakas, L. A.} and {Moutard, T.} and {M{\"u}ller, T.} and {Munari, E.} and {Murphree, G.} and {Murray, C.} and {Murray, N.} and {Musi, P.} and {Nadathur, S.} and {Nagam, B. C.} and {Nagao, T.} and {Naidoo, K.} and {Nakajima, R.} and {Nally, C.} and {Natoli, P.} and {Navarro-Alsina, A.} and {Navarro Girones, D.} and {Neissner, C.} and {Nersesian, A.} and {Nesseris, S.} and {Nguyen-Kim, H. N.} and {Nicastro, L.} and {Nichol, R. C.} and {Nielbock, M.} and {Niemi, S.-M.} and {Nieto, S.} and {Nilsson, K.} and {Noller, J.} and {Norberg, P.} and {Nouri-Zonoz, A.} and {Ntelis, P.} and {Nucita, A. A.} and {Nugent, P.} and {Nunes, N. J.} and {Nutma, T.} and {Ocampo, I.} and {Odier, J.} and {Oesch, P. A.} and {Oguri, M.} and {Magalhaes Oliveira, D.} and {Onoue, M.} and {Oosterbroek, T.} and {Oppizzi, F.} and {Ordenovic, C.} and {Osato, K.} and {Pacaud, F.} and {Pace, F.} and {Padilla, C.} and {Paech, K.} and {Pagano, L.} and {Page, M. J.} and {Palazzi, E.} and {Paltani, S.} and {Pamuk, S.} and {Pandolfi, S.} and {Paoletti, D.} and {Paolillo, M.} and {Papaderos, P.} and {Pardede, K.} and {Parimbelli, G.} and {Parmar, A.} and {Partmann, C.} and {Pasian, F.} and {Passalacqua, F.} and {Paterson, K.} and {Patrizii, L.} and {Pattison, C.} and {Paulino-Afonso, A.} and {Paviot, R.} and {Peacock, J. A.} and {Pearce, F. R.} and {Pedersen, K.} and {Peel, A.} and {Peletier, R. F.} and {Pellejero Ibanez, M.} and {Pello, R.} and {Penny, M. T.} and {Percival, W. J.} and {Perez-Garrido, A.} and {Perotto, L.} and {Pettorino, V.} and {Pezzotta, A.} and {Pezzuto, S.} and {Philippon, A.} and {Pierre, M.} and {Piersanti, O.} and {Pietroni, M.} and {Piga, L.} and {Pilo, L.} and {Pires, S.} and {Pisani, A.} and {Pizzella, A.} and {Pizzuti, L.} and {Plana, C.} and {Polenta, G.} and {Pollack, J. E.} and {Poncet, M.} and {P{\"o}ntinen, M.} and {Pool, P.} and {Popa, L. A.} and {Popa, V.} and {Popp, J.} and {Porciani, C.} and {Porth, L.} and {Potter, D.} and {Poulain, M.} and {Pourtsidou, A.} and {Pozzetti, L.} and {Prandoni, I.} and {Pratt, G. W.} and {Prezelus, S.} and {Prieto, E.} and {Pugno, A.} and {Quai, S.} and {Quilley, L.} and {Racca, G. D.} and {Raccanelli, A.} and {R{\'a}cz, G.} and {Radinovi{\'c}, S.} and {Radovich, M.} and {Ragagnin, A.} and {Ragnit, U.} and {Raison, F.} and {Ramos-Chernenko, N.} and {Ranc, C.} and {Rasera, Y.} and {Raylet, N.} and {Rebolo, R.} and {Refregier, A.} and {Reimberg, P.} and {Reiprich, T. H.} and {Renk, F.} and {Renzi, A.} and {Retre, J.} and {Revaz, Y.} and {Reyl{\'e}, C.} and {Reynolds, L.} and {Rhodes, J.} and {Ricci, F.} and {Ricci, M.} and {Riccio, G.} and {Ricken, S. O.} and {Rissanen, S.} and {Risso, I.} and {Rix, H.-W.} and {Robin, A. C.} and {Rocca-Volmerange, B.} and {Rocci, P.-F.} and {Rodenhuis, M.} and {Rodighiero, G.} and {Rodriguez Monroy, M.} and {Rollins, R. P.} and {Romanello, M.} and {Roman, J.} and {Romelli, E.} and {Romero-Gomez, M.} and {Roncarelli, M.} and {Rosati, P.} and {Rosset, C.} and {Rossetti, E.} and {Roster, W.} and {Rottgering, H. J. A.} and {Rozas-Fern{\'a}ndez, A.} and {Ruane, K.} and {Rubino-Martin, J. A.} and {Rudolph, A.} and {Ruppin, F.} and {Rusholme, B.} and {Sacquegna, S.} and {S{\'a}ez-Casares, I.} and {Saga, S.} and {Saglia, R.} and {Sahl{\'e}n, M.} and {Saifollahi, T.} and {Sakr, Z.} and {Salvalaggio, J.} and {Salvaterra, R.} and {Salvati, L.} and {Salvato, M.} and {Salvignol, J.-C.} and {S{\'a}nchez, A. G.} and {Sanchez, E.} and {Sanders, D. B.} and {Sapone, D.} and {Saponara, M.} and {Sarpa, E.} and {Sarron, F.} and {Sartori, S.} and {Sartoris, B.} and {Sassolas, B.} and {Sauniere, L.} and {Sauvage, M.} and {Sawicki, M.} and {Scaramella, R.} and {Scarlata, C.} and {Scharr{\'e}, L.} and {Schaye, J.} and {Schewtschenko, J. A.} and {Schindler, J.-T.} and {Schinnerer, E.} and {Schirmer, M.} and {Schmidt, F.} and {Schmidt, F.} and {Schmidt, M.} and {Schneider, A.} and {Schneider, M.} and {Schneider, P.} and {Sch{\"o}neberg, N.} and {Schrabback, T.} and {Schultheis, M.} and {Schulz, S.} and {Schuster, N.} and {Schwartz, J.} and {Sciotti, D.} and {Scodeggio, M.} and {Scognamiglio, D.} and {Scott, D.} and {Scottez, V.} and {Secroun, A.} and {Sefusatti, E.} and {Seidel, G.} and {Seiffert, M.} and {Sellentin, E.} and {Selwood, M.} and {Semboloni, E.} and {Sereno, M.} and {Serjeant, S.} and {Serrano, S.} and {Setnikar, G.} and {Shankar, F.} and {Sharples, R. M.} and {Short, A.} and {Shulevski, A.} and {Shuntov, M.} and {Sias, M.} and {Sikkema, G.} and {Silvestri, A.} and {Simon, P.} and {Sirignano, C.} and {Sirri, G.} and {Skottfelt, J.} and {Slezak, E.} and {Sluse, D.} and {Smith, G. P.} and {Smith, L. C.} and {Smith, R. E.} and {Smit, S. J. A.} and {Soldano, F.} and {Solheim, B. G. B.} and {Sorce, J. G.} and {Sorrenti, F.} and {Soubrie, E.} and {Spinoglio, L.} and {Spurio Mancini, A.} and {Stadel, J.} and {Stagnaro, L.} and {Stanco, L.} and {Stanford, S. A.} and {Starck, J.-L.} and {Stassi, P.} and {Steinwagner, J.} and {Stern, D.} and {Stone, C.} and {Strada, P.} and {Strafella, F.} and {Stramaccioni, D.} and {Surace, C.} and {Sureau, F.} and {Suyu, S. H.} and {Swindells, I.} and {Szafraniec, M.} and {Szapudi, I.} and {Taamoli, S.} and {Talia, M.} and {Tallada-Cresp{\'\i}, P.} and {Tanidis, K.} and {Tao, C.} and {Tarr{\'\i}o, P.} and {Tavagnacco, D.} and {Taylor, A. N.} and {Taylor, J. E.} and {Taylor, P. L.} and {Teixeira, E. M.} and {Tenti, M.} and {Teodoro Idiago, P.} and {Teplitz, H. I.} and {Tereno, I.} and {Tessore, N.} and {Testa, V.} and {Testera, G.} and {Tewes, M.} and {Teyssier, R.} and {Theret, N.} and {Thizy, C.} and {Thomas, P. D.} and {Toba, Y.} and {Toft, S.} and {Toledo-Moreo, R.} and {Tolstoy, E.} and {Tommasi, E.} and {Torbaniuk, O.} and {Torradeflot, F.} and {Tortora, C.} and {Tosi, S.} and {Tosti, S.} and {Trifoglio, M.} and {Troja, A.} and {Trombetti, T.} and {Tronconi, A.} and {Tsedrik, M.} and {Tsyganov, A.} and {Tucci, M.} and {Tutusaus, I.} and {Uhlemann, C.} and {Ulivi, L.} and {Urbano, M.} and {Vacher, L.} and {Vaillon, L.} and {Valageas, P.} and {Valdes, I.} and {Valentijn, E. A.} and {Valenziano, L.} and {Valieri, C.} and {Valiviita, J.} and {Van den Broeck, M.} and {Vassallo, T.} and {Vavrek, R.} and {Vega-Ferrero, J.} and {Venemans, B.} and {Venhola, A.} and {Ventura, S.} and {Verdoes Kleijn, G.} and {Vergani, D.} and {Verma, A.} and {Vernizzi, F.} and {Veropalumbo, A.} and {Verza, G.} and {Vescovi, C.} and {Vibert, D.} and {Viel, M.} and {Vielzeuf, P.} and {Viglione, C.} and {Viitanen, A.} and {Villaescusa-Navarro, F.} and {Vinciguerra, S.} and {Visticot, F.} and {Voggel, K.} and {von Wietersheim-Kramsta, M.} and {Vriend, W. J.} and {Wachter, S.} and {Walmsley, M.} and {Walth, G.} and {Walton, D. M.} and {Walton, N. A.} and {Wander, M.} and {Wang, L.} and {Wang, Y.} and {Weaver, J. R.} and {Weller, J.} and {Wetzstein, M.} and {Whalen, D. J.} and {Whittam, I. H.} and {Widmer, A.} and {Wiesmann, M.} and {Wilde, J.} and {Williams, O. R.} and {Winther, H.-A.} and {Wittje, A.} and {Wong, J. H. W.} and {Wright, A. H.} and {Yankelevich, V.} and {Yeung, H. W.} and {Yoon, M.} and {Youles, S.} and {Yung, L. Y. A.} and {Zacchei, A.} and {Zalesky, L.} and {Zamorani, G.} and {Zamorano Vitorelli, A.} and {Zanoni Marc, M.} and {Zennaro, M.} and {Zerbi, F. M.} and {Zinchenko, I. A.} and {Zoubian, J.} and {Zucca, E.} and {Zumalacarregui, M.}},
	doi = {10.1051/0004-6361/202450810},
	journal = {A\&A},
	pages = {A1},
	title = {Euclid - I. Overview of the Euclid mission},
	volume = 697,
	year = 2025}

@article{Veropalumbo_2021,
	author = {Veropalumbo, Alfonso and S{\'a}ez Casares, I{\~n}igo and Branchini, Enzo and Granett, Benjamin R and Guzzo, Luigi and Marulli, Federico and Moresco, Michele and Moscardini, Lauro and Pezzotta, Andrea and de la Torre, Sylvain},
	doi = {10.1093/mnras/stab2205},
	issn = {0035-8711},
	journal = {MNRAS},
	month = {07},
	number = {1},
	pages = {1184-1201},
	title = {A joint 2- and 3-point clustering analysis of the VIPERS PDR2 catalogue at $z \sim 1$: breaking the degeneracy of cosmological parameters},
	volume = {507},
	year = {2021}}

@article{Jeffrey_2020,
	author = {Jeffrey, Niall and Alsing, Justin and Lanusse, Fran{\c c}ois},
	doi = {10.1093/mnras/staa3594},
	journal = {MNRAS},
	month = {11},
	number = {1},
	pages = {954-969},
	title = {Likelihood-free inference with neural compression of DES SV weak lensing map statistics},
	volume = {501},
	year = {2020}}

@inproceedings{ioffe2015batch,
	author = {Ioffe, Sergey and Szegedy, Christian},
	booktitle = {International conference on machine learning},
	organization = {pmlr},
	pages = {448--456},
	title = {Batch normalization: Accelerating deep network training by reducing internal covariate shift},
	year = {2015}}

@article{Cranmer_2020,
	author = {{Cranmer}, Kyle and {Brehmer}, Johann and {Louppe}, Gilles},
	doi = {10.1073/pnas.1912789117},
	journal = {Proceedings of the National Academy of Science},
	month = dec,
	number = {48},
	pages = {30055-30062},
	title = {{The frontier of simulation-based inference}},
	volume = {117},
	year = 2020}

@article{Ho_2024,
	author = {{Ho}, Matthew and {Bartlett}, Deaglan J. and {Chartier}, Nicolas and {Cuesta-Lazaro}, Carolina and {Ding}, Simon and {Lapel}, Axel and {Lemos}, Pablo and {Lovell}, Christopher C. and {Makinen}, T. Lucas and {Modi}, Chirag and {Pandya}, Viraj and {Pandey}, Shivam and {Perez}, Lucia A. and {Wandelt}, Benjamin and {Bryan}, Greg L.},
	doi = {10.33232/001c.120559},
	eid = {54},
	journal = {The Open Journal of Astrophysics},
	month = jul,
	pages = {54},
	title = {{LtU-ILI: An All-in-One Framework for Implicit Inference in Astrophysics and Cosmology}},
	volume = {7},
	year = 2024}

@article{Marin_2012,
	author = {{Marin}, Jean-Michel and {Pudlo}, Pierre and {Robert}, Christian P. and {Ryder}, Robin},
	doi = {10.1007/s11222-011-9288-2},
	journal = {Statistics and Computing},
	pages = {1167-1180},
	title = {{Approximate Bayesian Computational methods}},
	volume = {22},
	year = {2012}}

@article{Maksimova_2021,
	author = {Maksimova, Nina A and Garrison, Lehman H and Eisenstein, Daniel J and Hadzhiyska, Boryana and Bose, Sownak and Satterthwaite, Thomas P},
	doi = {10.1093/mnras/stab2484},
	issn = {0035-8711},
	journal = {MNRAS},
	month = {09},
	number = {3},
	pages = {4017-4037},
	title = {AbacusSummit: a massive set of high-accuracy, high-resolution N-body simulations},
	volume = {508},
	year = {2021}}

@article{Leclercq_2021,
	author = {Leclercq, Florent and Heavens, Alan},
	doi = {10.1093/mnrasl/slab081},
	issn = {1745-3925},
	journal = {Monthly Notices of the Royal Astronomical Society: Letters},
	month = {07},
	number = {1},
	pages = {L85-L90},
	title = {On the accuracy and precision of correlation functions and field-level inference in cosmology},
	volume = {506},
	year = {2021}}

@article{Borgani_2001,
	author = {Borgani, Stefano and Guzzo, Luigi},
	day = {01},
	doi = {10.1038/35051000},
	issn = {1476-4687},
	journal = {Nature},
	month = {Jan},
	number = {6816},
	pages = {39-45},
	title = {X-ray clusters of galaxies as tracers of structure in the Universe},
	url = {https://doi.org/10.1038/35051000},
	volume = {409},
	year = {2001}}

\end{document}